\documentclass[
  nojss]{jss}

\usepackage[utf8]{inputenc}

\providecommand{\tightlist}{%
  \setlength{\itemsep}{0pt}\setlength{\parskip}{0pt}}

\author{
Nicolas Bennett\footnotemark[1]\\ETH Zürich \And Drago
Plečko\footnotemark[1]\footnotetext[1]{These authors contributed equally.}\\ETH
Zürich \And Ida-Fong Ukor\\Monash Health \AND Nicolai
Meinshausen\\ETH Zürich \And Peter Bühlmann\\ETH Zürich
}
\title{\pkg{ricu}: \proglang{R}'s Interface to Intensive Care Data}

\Plainauthor{Nicolas Bennett, Drago Plečko, Ida-Fong
Ukor, Nicolai Meinshausen, Peter Bühlmann}
\Plaintitle{ricu: R's Interface to Intensive Care Data}
\Shorttitle{\pkg{ricu}: R meets ICU data}

\Abstract{
Providing computational infrastructure for handling diverse intensive
care unit (ICU) datasets, the \proglang{R} package \pkg{ricu} enables
writing dataset-agnostic analysis code, thereby facilitating
multi-center training and validation of machine learning models. The
package is designed with an emphasis on extensibility both to new
datasets as well as clinical data concepts, and currently supports the
loading of around 100 patient variables corresponding to a total of
319,402 ICU admissions from 4 data sources collected in Europe and the
United States. By allowing for the addition of user-specified medical
concepts and data sources the aim of \pkg{ricu} is to foster robust,
data-based intensive care research, allowing the user to externally
validate their method or conclusion with relative ease, and in turn
facilitating reproducible and therefore transparent work in this field.
}

\Plainkeywords{electronic health records, computational
physiology, critical care medicine}

\Address{
    Nicolas Bennett\footnotemark[1]\\
    ETH Zürich\\
    Seminar for Statistics\\
Rämistrasse 101\\
CH-8092 Zurich\\
  E-mail: \email{nicolas.bennett@stat.math.ethz.ch}\\
  
      Drago
Plečko\footnotemark[1]\footnotetext[1]{These authors contributed equally.}\\
    ETH Zürich\\
    Seminar for Statistics\\
Rämistrasse 101\\
CH-8092 Zürich\\
  E-mail: \email{drago.plecko@stat.math.ethz.ch}\\
  
      Ida-Fong Ukor\\
    Monash Health\\
    Department of Anaesthesiology and Perioperative Medicine\\
246 Clayton Road\\
Clayton VIC 3168\\
  E-mail: \email{ida-fong.ukor@monashhealth.org}\\
  
      Nicolai Meinshausen\\
    ETH Zürich\\
    Seminar for Statistics\\
Rämistrasse 101\\
CH-8092 Zürich\\
  E-mail: \email{meinshausen@stat.math.ethz.ch}\\
  
      Peter Bühlmann\\
    ETH Zürich\\
    Seminar for Statistics\\
Rämistrasse 101\\
CH-8092 Zürich\\
  E-mail: \email{peter.buehlmann@stat.math.ethz.ch}\\
  
  }

\usepackage{amsmath} \usepackage{booktabs} \usepackage{makecell} \usepackage{threeparttablex}

\begin{document}

\maketitle

\renewcommand*{\thefootnote}{\fnsymbol{footnote}}
\footnotetext{$^{*}$These authors contributed equally.}
\renewcommand*{\thefootnote}{\arabic{footnote}}

\hypertarget{introduction}{%
\section{Introduction}\label{introduction}}

Collection of electronic health records has seen a significant rise in
recent years \citep{evans2016}, opening up opportunities and providing
the grounds for a large body of data-driven research oriented towards
helping clinicians in decision-making and therefore improving patient
care and health outcomes \citep{jiang2017}.

One example of a problem that has received much attention from the
machine learning community is early prediction of sepsis in ICU
\citep{desautels2016, nemati2018, futoma2017, kam2017}. Interestingly,
there is evidence that a large proportion of the publications are based
on the same dataset \citep{fleuren2019}, the Medical Information Mart
for Intensive Care III \citep[MIMIC-III;][]{johnson2016}, which shows a
systematic lack of external validation. Part of this problem might well
be the need for computational infrastructure handling multiple datasets.
The MIMIC-III dataset consists of 26 different tables containing about
20GB of data. While much work and care has gone into data preprocessing
in order to provide a self-contained ready -to-use data resource with
MIMIC-III, seemingly simple tasks such as computing a sepsis-related
organ failure assessment (SOFA) score \citep{vincent1996} remains a
non-trivial effort\footnote{There is considerable heterogeneity in
  number of patients satisfying the Sepsis-3 criterion
  \citep[of which SOFA provides a major component;][]{singer2016} among
  studies investigating MIMIC-III. Reported Sepsis-3 prevalence ranges
  from 11.3\% \citep{desautels2016}, over 23.9\% \citep{nemati2018} and
  25.4\% \citep{wang2018}, up to 49.1\% \citep{johnson2018}. While some
  of this variation may be explained by differing patient inclusion
  criteria, diversity in label implementation must also contribute
  significantly.}. This is only exacerbated when aiming to co-integrate
multiple different datasets of this form, spanning hospitals and even
countries, in order to capture effects of differing practice and
demographics.

The aim of the
\href{https://cran.r-project.org/package=ricu}{\pkg{ricu}} package is to
provide computational infrastructure allowing users to investigate
complex research questions in the context of critical care medicine as
easily as possible by providing a unified interface to a heterogeneous
set of data sources. The package enables users to write dataset-agnostic
code which can simplify implementation and shorten the time necessary
for prototyping code querying different datasets. In its current form,
the package handles four large-scale, publicly available intensive care
databases out of the box: MIMIC-III from the Beth Israel Deaconess
Medical Center in Boston, Massachusetts \citep{johnson2016}, the eICU
Collaborative Research Database \citep{pollard2018}, containing data
collected from 208 hospitals across the United States, the High Time
Resolution ICU Dataset (HiRID) from the Department of Intensive Care
Medicine of the Bern University Hospital, Switzerland \citep{faltys2021}
and AmsterdamUMCdb from the Amsterdam University Medical Center
\citep{thoral2021}. Furthermore, \pkg{ricu} was designed with
extensibility in mind such that adding further public and/or private
user-provided datasets is possible. Being implemented in \proglang{R}, a
programming language popular among statisticians and data analysts, it
is our hope to contribute to accessible and reproducible research by
using a familiar environment and requiring only few system dependencies,
thereby simplifying setup considerably.

To our knowledge, infrastructure that provides a common interface to
multiple such datasets is a novel contribution. While there have been
efforts \citep{adibuzzaman2016, wang2020} attempting to abstract away
some specifics of a dataset, these have so far exclusively focused on
MIMIC-III, the most popular of public ICU datsets and have not been
designed with dataset interoperability in mind.

Given the somewhat narrow focus of the targeted datasets, combined with
the fact that in some cases data is even extracted from identical
patient care management systems, it may come as a surprise as to how
heterogeneous the resulting datasets are. In MIMIC-III and HiRID, for
example, time-stamps are reported as absolute times (albeit randomly
shifted due to data privacy concerns), whereas eICU and AUMC use
relative times (with origins being admission times). Another example,
involves different types of patient identifiers and their use among
datasets. Common to all is the notion of an ICU admission ID, but apart
from that, the amount of available information varies: While ICU (and
hospital) readmissions for a given patient can be identified in some,
this is not possible in other datasets. Furthermore, use of identifier
systems might not be consistent over tables. In MIMIC-III, for example,
some tables refer to ICU stay IDs while others use hospital stay IDs,
which slightly complicates data retrieval for a given ID system.
Additionally, table layouts vary (\emph{long} versus \emph{wide} data
arrangement) and data organization in general is far from consistent
over datasets.

\hypertarget{quick-start-guide}{%
\section{Quick start guide}\label{quick-start-guide}}

The following list gives a quick outline of the steps required for
setting up and starting to use \pkg{ricu}, alongside some section
references on where to find further details. A more comprehensive
version of this overview is available as a
\href{https://CRAN.R-project.org/package=ricu/vignettes/ricu.html}{separate
vignette}.

\begin{enumerate}
\def\labelenumi{\arabic{enumi}.}
\item
  Package installation:

  \begin{itemize}
  \item
    the latest release of \pkg{ricu} can be installed from CRAN as
    \texttt{install.packages("ricu")}
  \item
    alternatively, the latest development version is available from
    \href{https://github.com/eth-mds/ricu}{Github} by running
    \texttt{remotes::install\_github("eth-mds/ricu")}
  \end{itemize}
\item
  Requesting access to datasets and data source setup:

  \begin{itemize}
  \item
    demo datasets can be set up by installing the data packages
    \texttt{mimic.demo} and/or \texttt{eicu.demo}, passing
    \texttt{"https://eth-mds.github.io/physionet-demo"} as
    \texttt{repos} argument to \texttt{install.packages()}
  \item
    the complete MIMIC-III, eICU and HiRID datasets can be accessed by
    setting up an account at
    \href{https://physionet.org/register}{PhysioNet}
  \item
    access to AUMCdb is available via the
    \href{https://amsterdammedicaldatascience.nl/\#amsterdamumcdb}{Amsterdam
    Medical Data Science Website}
  \item
    the obtained credentials can be configured for PhysioNet datasets by
    setting environment variables \texttt{RICU\_PHYSIONET\_USER} and
    \texttt{RICU\_PHYSIONET\_PASS}, while the download token for AUMCdb
    can be set as \texttt{RICU\_AUMC\_TOKEN}
  \item
    datasets are downloaded and set up either automatically upon the
    first access attempt or manually by running
    \texttt{setup\_data\_src()}; the environment variable
    \texttt{RICU\_DATA\_PATH} can be set to control data location
  \item
    dataset availability can be queried by calling
    \texttt{src\_data\_avail()}
  \end{itemize}

  A more detailed description of the datasets and the setup process is
  given in Section \ref{data-sources}, with Section
  \ref{ready-to-use-datasets} providing an overview of each of the 4
  supported datasets and Section \ref{implementation-details}
  elaborating on how datasets are represented in code.
\item
  Loading of data corresponding to clinical concepts using
  \texttt{load\_concepts()}:

  \begin{itemize}
  \item
    currently, over 100 data concepts are available for the 4 supported
    datasets (see
    \texttt{concept\_availability()}/\texttt{explain\_dictionary()} for
    names, availability etc.)
  \item
    both concepts and data sources can be specified as strings; for
    example, glucose and age data can be loaded from MIMIC-III as
    \texttt{load\_concepts(c("age",\ "glu"),\ "mimic")}
  \end{itemize}

  Section \ref{data-concepts} goes into more detail on how data concepts
  are represented within \pkg{ricu} and an overview of the
  pre-configured concepts is available from Section
  \ref{clinical-concepts}.
\item
  Extending the concept dictionary:

  \begin{itemize}
  \item
    data concepts can be specified in code using the constructors
    \texttt{concept()}/\texttt{item()} or
    \texttt{new\_concept()}/\texttt{new\_item()}
  \item
    for session persistence, data concepts can also be specified as JSON
    formatted objects
  \item
    JSON-based concept dictionaries can either extend or replace others
    and they can be pointed to by setting the environment variable
    \texttt{RICU\_CONFIG\_PATH}
  \end{itemize}

  The JSON format used to encode data concepts is discussed in more
  detail in Section \ref{concept-specification}.
\item
  Adding new datasets:

  \begin{itemize}
  \item
    a JSON-based dataset configuration file is required, from which the
    configuration objects described in Section
    \ref{implementation-details} are created
  \item
    in order for concepts to be available from the new dataset, the
    dictionary requires extension by adding new data items
  \end{itemize}

  Some further information about adding a custom dataset is available
  from Section \ref{adding-external-datasets}, albeit not in much
  detail. Some code used when AUMCdb was not yet fully integrated with
  \pkg{ricu} is available from
  \href{https://github.com/eth-mds/aumc}{Github}.
\end{enumerate}

The final section (\ref{examples}) shows briefly how \pkg{ricu} could be
used in practice to address clinical questions by presenting two small
examples.

\hypertarget{data-sources}{%
\section{Data sources}\label{data-sources}}

In order to make data available from different data sources, \pkg{ricu}
provides abstractions using JSON-formatted configuration files and a set
of S3 classes with associated S3 generic functions. This system is
designed with extensibility in mind, allowing for incorporation of a
wide variety of datasets. Provisions for several large-scale publicly
available datasets in terms of required configuration information
alongside class-specific implementations of the needed S3 generic
functions are part of \pkg{ricu}, opening up access to these datasets.
Data itself, however, is not part of \pkg{ricu} but rather can be
downloaded from the Internet using tools provided by \pkg{ricu}. While
the datasets are publicly available, access has to be granted by the
dataset creators individually. Three datasets, MIMIC-III, eICU and HiRID
are hosted on PhysioNet \citep{goldberger2000}, access to which requires
an \href{https://physionet.org/register/}{account}, while the fourth,
AmsterdamUMCdb is currently distributed via a separate platform,
requiring a
\href{https://amsterdammedicaldatascience.nl/\#amsterdamumcdb}{download
link}.

For both MIMIC-III and eICU, small subsets of data are available as demo
datasets that do not require credentialed access to PhysioNet. As the
terms for distribution of these demo datasets are less restrictive, they
can be made available as data packages \pkg{mimic.demo} and
\pkg{eicu.demo}. Due to size constraints, however they are not available
via CRAN, but can be installed from Github as

\begin{CodeChunk}
\begin{CodeInput}
R> install.packages(
+   c("mimic.demo", "eicu.demo"),
+   repos = "https://eth-mds.github.io/physionet-demo"
+ )
\end{CodeInput}
\end{CodeChunk}

Provisions for datasets configured to be attached during package loading
are made irrespective of whether data is actually available. Upon access
of an incomplete dataset, the user is asked for permission to download
in interactive sessions and an error is thrown otherwise. Credentials
can either be provided as environment variables
(\texttt{RICU\_PHYSIONET\_USER} and \texttt{RICU\_PHYSIONET\_PASS} for
access to PhysioNet data, as well as \texttt{RICU\_AUMC\_TOKEN} for
AmsterdamUMCdb) and if the corresponding variables are unset, user input
is again required in interactive sessions. For non-interactive sessions,
functionality is exported such that data can be downloaded and set up
ahead of first access (see \texttt{?setup\_src\_data}).

\hypertarget{ready-to-use-datasets}{%
\subsection{Ready to use datasets}\label{ready-to-use-datasets}}

Contingent on being granted access by the data owners, several
large-scale ICU datasets collected from multiple hospitals in the US and
Europe can be set up for access using \pkg{ricu} with minimal user
effort. Download requires a stable Internet connection, as well as 50 to
100 GB of temporary disk storage for unpacking and preparing the data
for efficient access. In terms of permanent storage, 5 to 10 GB per
dataset are required, while memory requirements permit importing (and
working with) even the largest tables using Laptop class hardware as
only subsets of rows are read at once.

The following paragraphs serve to give quick introductions to the
included datasets to offer some guidance, outlining some strengths and
weaknesses of each of the datasets. Especially the PhysioNet datasets
\href{https://mimic.physionet.org/about/mimic/}{MIMIC-III} and
\href{https://eicu-crd.mit.edu/about/eicu/}{eICU} offer good
documentation on the respective websites. This section is concluded with
a table summarizing similarities and differences among the datasets,
outlined in the following paragraphs (see table \ref{tab:datasets}).

\hypertarget{mimic-iii}{%
\subsubsection{MIMIC-III}\label{mimic-iii}}

The \href{https://physionet.org/content/mimiciii/1.4/}{Medical
Information Mart for Intensive Care III (MIMIC-III)} represents the
third iteration of the arguably most influential initiative for
collecting and providing to the public large-scale ICU data\footnote{The
  initial MIMIC (at the time short for Multi-parameter Intelligent
  Monitoring for Intensive Care) data release dates back 20 years and
  contained data on roughly 100 patients recorded from patient monitors
  in the medical, surgical, and cardiac intensive care units of Boston's
  Beth Israel Hospital during the years 1992-1999 \citep{moody1996}.
  Significantly broadened in scope, MIMIC-II was released 10 years
  after, now including data on almost 27,000 adult hospital admissions
  collected from ICUs of Beth Israel Deaconess Medical Center (BIDMC)
  from 2001 to 2008 \citep{lee2011}. Following MIMIC-III, release of
  MIMIC-IV is imminent with a first development version having been
  released in summer 2020. This iteration of MIMIC too is planned to be
  included with \pkg{ricu} as soon a first stable version is released.}.
The dataset comprises de-identified health related data of roughly
46,000 patients admitted to critical care units of BIDMC during the
years 2001-2012. Amounting to just over 61,000 individual ICU admission,
data is available on demographics, routine vital sign measurements (at
approximately 1 hour resolution), laboratory tests, medication, as well
as critical care procedures, organized as a 26-table relational
structure.

\begin{CodeChunk}
\begin{CodeInput}
R> mimic
\end{CodeInput}
\begin{CodeOutput}
<mimic_env[26]>
        admissions            callout         caregivers        chartevents 
     [58,976 x 19]      [34,499 x 24]        [7,567 x 4] [330,712,483 x 15] 
         cptevents              d_cpt    d_icd_diagnoses   d_icd_procedures 
    [573,146 x 12]          [134 x 9]       [14,567 x 4]        [3,882 x 4] 
           d_items         d_labitems     datetimeevents      diagnoses_icd 
     [12,487 x 10]          [753 x 6]   [4,485,937 x 14]      [651,047 x 5] 
          drgcodes           icustays     inputevents_cv     inputevents_mv 
     [125,557 x 8]      [61,532 x 12]  [17,527,935 x 22]   [3,618,991 x 31] 
         labevents microbiologyevents         noteevents       outputevents 
  [27,854,055 x 9]     [631,726 x 16]   [2,083,180 x 11]   [4,349,218 x 13] 
          patients      prescriptions procedureevents_mv     procedures_icd 
      [46,520 x 8]   [4,156,450 x 19]     [258,066 x 25]      [240,095 x 5] 
          services          transfers 
      [73,343 x 6]     [261,897 x 13] 
\end{CodeOutput}
\end{CodeChunk}

One thing of note from a data-organizational perspective is that a
change in electronic health care systems occurred in 2008. Owing to
this, roughly 48,000 ICU admissions spanning the years 2001 though 2008
are documented using the CareVue system, while for 2008 and onwards,
data was extracted from the MetaVision system. Item identifiers differ
between the two systems, requiring queries to consider both ID mappings
(heart rate for example being available both as \texttt{itemid} number
\texttt{211} for CareVue and \texttt{220045} for MetaVision) as does
documentation of infusions and other procedures that are considered as
input events (c.f. \texttt{inputevents\_cv} and \texttt{inputevents\_mv}
tables). Especially with respect to such input event data, MetaVision
data generally is of superior quality.

In terms of patient identifiers, MIMIC-III allows for identifying both
individual patients (\texttt{subject\_id}) across hospital admissions
(\texttt{hadm\_id}) and for connecting ICU (re-)admissions
(\texttt{icustay\_id}) to hospital admissions. Using the respective
one-to-many relationships, \pkg{ricu} can retrieve patient data using
any of the above IDs, irrespective of how the raw data is organized.

\hypertarget{eicu}{%
\subsubsection{eICU}\label{eicu}}

Unlike the single-center focus of other datasets, the
\href{https://physionet.org/content/eicu-crd/2.0/}{eICU Collaborative
Research Database} constitutes an amalgamation of data from critical
care units of over 200 hospitals throughout the continental United
States. Large-scale data collected via the Philips eICU program which
provides telehealth infrastructure for intensive care units, is
available from the Philips eICU Research Institute (eRI), albeit neither
publicly nor freely. Only data corresponding to roughly 200,000 ICU
admissions, sampled from a larger population of over 3 million ICU
admissions and stratified by hospital, is being made available via
PhysioNet. Patients with discharge dates in 2014 or 2015 were
considered, with stays in low acuity units being removed.

\begin{CodeChunk}
\begin{CodeInput}
R> eicu
\end{CodeInput}
\begin{CodeOutput}
<eicu_env[31]>
            admissiondrug               admissiondx                   allergy 
           [874,920 x 14]             [626,858 x 6]            [251,949 x 13] 
             apacheapsvar       apachepatientresult             apachepredvar 
           [171,177 x 26]            [297,064 x 23]            [171,177 x 51] 
     careplancareprovider               careplaneol           careplangeneral 
            [502,765 x 8]               [1,433 x 5]           [3,115,018 x 6] 
             careplangoal careplaninfectiousdisease                 customlab 
            [504,139 x 7]               [8,056 x 8]               [1,082 x 7] 
                diagnosis                  hospital              infusiondrug 
          [2,710,672 x 7]                 [208 x 4]           [4,803,719 x 9] 
             intakeoutput                       lab                medication 
        [12,030,289 x 12]         [39,132,531 x 10]          [7,301,853 x 15] 
                 microlab                      note           nurseassessment 
             [16,996 x 7]           [2,254,179 x 8]          [15,602,498 x 8] 
                nursecare             nursecharting               pasthistory 
          [8,311,132 x 8]         [151,604,232 x 8]           [1,149,180 x 8] 
                  patient              physicalexam           respiratorycare 
           [200,859 x 29]           [9,212,316 x 6]            [865,381 x 34] 
      respiratorycharting                 treatment            vitalaperiodic 
         [20,168,176 x 7]           [3,688,745 x 5]         [25,075,074 x 13] 
            vitalperiodic 
       [146,671,642 x 19] 
\end{CodeOutput}
\end{CodeChunk}

The data is organized into 31 tables and includes patient demographics,
routine vital signs, laboratory measurements, medication
administrations, admission diagnoses, as well as treatment information.
Owing to the wide range of hospitals participating in this data
collection initiative, spanning small, rural, non-teaching health
centers with fewer than 100 beds to large teaching hospitals with an
excess of 500 beds, data availability varies. Even if data was being
recorded at the bedside it might end up missing from the eICU dataset
due to technical limitations of the collection process. As for patient
identifiers, while it is possible to link ICU admissions corresponding
to the same hospital stay, it is not possible to identify patients
across hospital stays.

Data resolution again varies considerably over included variables. The
\texttt{vitalperiodic} table stands out as one of the few examples of a
\emph{wide} table organization (laying out variables as columns), as
opposed to the \emph{long} presentation (following an
entity--attribute--value) of most other tables containing patient
measurement data. The average time step in \texttt{vitalperiodic} is
around 5 minutes, but data missingness ranges from around 1\% for heart
rate and pulse oximetry to around 80-90\% for blood pressure
measurements, therefore giving approximately hourly resolution for such
variables.

\hypertarget{hirid}{%
\subsubsection{HiRID}\label{hirid}}

Developed for early prediction of circulatory failure
\citep{hyland2020}, the
\href{https://physionet.org/content/hirid/1.0/}{High Time Resolution ICU
Dataset (HiRID)} contains data on almost 34,000 admissions to the
Department of Intensive Care Medicine of the Bern University Hospital,
Switzerland, an interdisciplinary 60-bed unit. Given the clear focus on
a concrete application during data collection, this dataset is the most
limited in terms of breadth of available information, which is also
reflected in a comparatively simple data layout comprising only 5
tables\footnote{The data is available in three states: as raw data and
  in preprocessed form, with preprocessed data being represented by two
  intermediary pipeline stages from \citep{hyland2020}. While \pkg{ricu}
  focuses exclusively on raw data, the \emph{merged} stage represents a
  selection of variables that were deemed most predictive for
  determining circulatory failure, which are then merged into 18
  meta-variables, representing different clinical concepts. Time stamps
  in \emph{merged} data are left unchanged, yielding irregular time
  series, whereas for the \emph{imputed} stage, data is down-sampled to
  a 5 minute grid and missing values are imputed using a scheme
  discussed in \citep{hyland2020}.}.

\begin{CodeChunk}
\begin{CodeInput}
R> hirid
\end{CodeInput}
\begin{CodeOutput}
<hirid_env[5]>
          general      observations           ordinal            pharma 
     [33,905 x 5] [776,921,131 x 8]          [72 x 3] [16,270,399 x 14] 
        variables 
        [712 x 5] 
\end{CodeOutput}
\end{CodeChunk}

Collected during the period of January 2008 through June 2016, roughly
700 distinct variables covering routine vital signs, diagnostic test
results and treatment parameters are available with variables monitored
at the bedside being recorded with two minute time resolution. In terms
of demographic information and patient identifier systems however, the
data is limited. It is not possible to identify ICU admissions
corresponding to individual patients and apart from patient age, sex,
weight and height, very little information is available to characterize
patients. There is no medical history, no admission diagnoses, only
in-ICU mortality information, no unstructured patient data and no
information on patient discharge. Furthermore, data on body fluid
sampling has been omitted, complicating for example the construction of
a Sepsis-3 label \citep{singer2016}.

\hypertarget{amsterdamumcdb}{%
\subsubsection{AmsterdamUMCdb}\label{amsterdamumcdb}}

As a second European dataset, also focusing on increased time-resolution
over the US datasets,
\href{https://amsterdammedicaldatascience.nl/\#amsterdamumcdb}{AmsterdamUMCdb}
has been made available in late 2019, containing data on over 23,000
intensive care unit and high dependency unit admissions of adult
patients during the years 2003 through 2016. The department of Intensive
Care at Amsterdam University Medical Center is a mixed medical-surgical
ICU with up to 32 bed ICU and 12 bed high dependency units with an
average of 1000-2000 yearly admissions. Covering middle ground between
the US datasets and HiRID in terms of breadth of included data, while
providing a maximal time-resolution of 1 minute, AmsterdamUMCdb
constitutes a well organized high quality ICU data resource organized
succinctly as a 7-table relational structure.

\begin{CodeChunk}
\begin{CodeInput}
R> aumc
\end{CodeInput}
\begin{CodeOutput}
<aumc_env[7]>
         admissions           drugitems       freetextitems           listitems 
      [23,106 x 19]    [4,907,269 x 31]      [651,248 x 11]   [30,744,065 x 11] 
       numericitems procedureorderitems        processitems 
 [977,625,612 x 15]     [2,188,626 x 8]       [256,715 x 6] 
\end{CodeOutput}
\end{CodeChunk}

A slightly different approach to data anonymization was chosen for this
dataset, yielding demographic information such as patient weight, height
and age only available as binned variables instead of raw numeric
values. Apart from this, there is information on patient origin,
mortality, admission diagnoses, as well as numerical measurements
including vital parameters, lab results, outputs from drains and
catheters, information on administered medication, and other medical
procedures. In terms of patient identifiers, it is possible to link ICU
admissions corresponding to the same individual, but it is not possible
to identify separate hospital admissions.

\begin{table}

\caption{\label{tab:datasets}Comparison of datasets supported by \pkg{ricu}, highlighting some of the major similarities and distinguishing features among the four data sources described in the preceding paragraphs. Values followed by parenthesized ranges represent medians and are accompanied by quartiles.}
\centering
\begin{threeparttable}
\begin{tabular}[t]{lllll}
\toprule
  & MIMIC & eICU & HiRID & AUMC\\
\midrule
Number of tables & 26 & 31 & 5 & 7\\
Disk storage [GB] & 7.58 & 6.49 & 4.52 & 13.65\\
Available concepts\textsuperscript{*} & 88 & 85 & 74 & 85\\
\addlinespace[0.3em]
\multicolumn{5}{l}{\textbf{Admission counts}}\\
\hspace{1em}ICU & 61,522 & 200,857 & 33,904 & 23,106\\
\hspace{1em}Hospital & 57,702 & 166,342 & - & -\\
\hspace{1em}Unique patients & 46,520 & - & - & 19,681\\
\addlinespace[0.3em]
\multicolumn{5}{l}{\textbf{Stay lengths [hr]}}\\
\hspace{1em}ICU stays & \makecell[l]{2.09\\(1.11 - 4.48)} & \makecell[l]{1.57\\(0.82 - 2.97)} & \makecell[l]{0.99\\(0.81 - 2.16)} & \makecell[l]{1.07\\(0.84 - 3.67)}\\
\hspace{1em}Hospital stays & \makecell[l]{6.57\\(3.80 - 11.86)} & \makecell[l]{5.05\\(2.71 - 9.03)} & - & -\\
\addlinespace[0.3em]
\multicolumn{5}{l}{\textbf{Frequency [1/hr]}}\\
\hspace{1em}\makecell[l]{Vital signs\\(heart rate)} & \makecell[l]{1.00\\(1.00 - 1.02)} & \makecell[l]{12.00\\(12.00 - 12.00)} & \makecell[l]{30.00\\(30.00 - 60.00)} & \makecell[l]{60.00\\(60.00 - 60.00)}\\
\hspace{1em}\makecell[l]{Lab tests\\(bilirubin)} & \makecell[l]{0.04\\(0.04 - 0.06)} & \makecell[l]{0.04\\(0.04 - 0.05)} & \makecell[l]{0.04\\(0.04 - 0.04)} & \makecell[l]{0.04\\(0.01 - 0.04)}\\
\bottomrule
\end{tabular}
\begin{tablenotes}
\item[*] These values represent the number of atomic concepts per data source. Additionally, 29 recursive concepts are available, which build on source-specific atomic concepts in a source-agnostic manner (see Section \ref{concept-specification} for details).
\end{tablenotes}
\end{threeparttable}
\end{table}

\hypertarget{implementation-details}{%
\subsection{Implementation details}\label{implementation-details}}

Every dataset is represented by an environment with class attributes and
associated metadata objects stored as object attributes to that
environment. Dataset environments all inherit from \texttt{src\_env} and
from any number of class names constructed from data source name(s) with
a suffix \texttt{\_env} attached. The environment representing
MIMIC-III, for example inherits from \texttt{src\_env} and
\texttt{mimic\_env}, while the corresponding demo dataset inherits from
\texttt{src\_env}, \texttt{mimic\_env} and \texttt{mimic\_demo\_env}.
These sub-classes are later used for tailoring the process of data
loading to particularities of individual datasets.

A \texttt{src\_env} contains an active binding per contained table,
which returns a \texttt{src\_tbl} object representing the requested
table. As is the case for \texttt{src\_env} objects, \texttt{src\_tbl}
objects inherit from additional classes such that certain per-dataset
behavior can be customized. The \texttt{admissions} table of the
MIMIC-III demo dataset for example inherits from
\texttt{mimic\_demo\_tbl} and \texttt{mimic\_tbl} (alongside classes
\texttt{src\_tbl} and \texttt{prt}).

\begin{CodeChunk}
\begin{CodeInput}
R> mimic_demo$admissions
\end{CodeInput}
\begin{CodeOutput}
# <mimic_tbl>: [129 x 19]
# ID options:  subject_id (patient) < hadm_id (hadm) < icustay_id (icustay)
# Defaults:    `admission_type` (val)
# Time vars:   `admittime`, `dischtime`, `deathtime`, `edregtime`, `edouttime`
    row_id subject_id hadm_id admittime           dischtime
     <int>      <int>   <int> <dttm>              <dttm>
  1  12258      10006  142345 2164-10-23 21:09:00 2164-11-01 17:15:00
  2  12263      10011  105331 2126-08-14 22:32:00 2126-08-28 18:59:00
  3  12265      10013  165520 2125-10-04 23:36:00 2125-10-07 15:13:00
  4  12269      10017  199207 2149-05-26 17:19:00 2149-06-03 18:42:00
  5  12270      10019  177759 2163-05-14 20:43:00 2163-05-15 12:00:00
...
125  41055      44083  198330 2112-05-28 15:45:00 2112-06-07 16:50:00
126  41070      44154  174245 2178-05-14 20:29:00 2178-05-15 09:45:00
127  41087      44212  163189 2123-11-24 14:14:00 2123-12-30 14:31:00
128  41090      44222  192189 2180-07-19 06:55:00 2180-07-20 13:00:00
129  41092      44228  103379 2170-12-15 03:14:00 2170-12-24 18:00:00
# ... with 119 more rows, and 14 more variables: deathtime <dttm>,
#   admission_type <chr>, admission_location <chr>, discharge_location <chr>,
#   insurance <chr>, language <chr>, religion <chr>, marital_status <chr>,
#   ethnicity <chr>, edregtime <dttm>, edouttime <dttm>, diagnosis <chr>,
#   hospital_expire_flag <int>, has_chartevents_data <int>
\end{CodeOutput}
\end{CodeChunk}

Powered by the \pkg{prt} \citep{bennett2021} package, \texttt{src\_tbl}
objects represent row-partitioned tabular data stored as multiple binary
files created by the \pkg{fst} \citep{klik2020} package. In addition to
standard subsetting, \texttt{prt} objects can be subsetted via the base
R S3 generic function \texttt{subset()} and using non-standard
evaluation:

\begin{CodeChunk}
\begin{CodeInput}
R> subset(mimic_demo$admissions, subject_id > 44000, language:ethnicity)
\end{CodeInput}
\begin{CodeOutput}
   language          religion marital_status              ethnicity
1:     ENGL          CATHOLIC         SINGLE                  WHITE
2:     ENGL          CATHOLIC         SINGLE                  WHITE
3:     ENGL          CATHOLIC         SINGLE                  WHITE
4:     ENGL PROTESTANT QUAKER        MARRIED                  WHITE
5:     ENGL      UNOBTAINABLE         SINGLE BLACK/AFRICAN AMERICAN
6:     ENGL          CATHOLIC         SINGLE                  WHITE
7:     ENGL     NOT SPECIFIED         SINGLE                  WHITE
\end{CodeOutput}
\end{CodeChunk}

This syntax makes it possible to read row-subsets of \emph{long} tables
into memory with little memory overhead. While terseness of such an API
does introduce potential ambiguity, this is mostly overcome by using the
tidy eval framework provided by \pkg{rlang} \citep{wickham2020}:

\begin{CodeChunk}
\begin{CodeInput}
R> subject_id <- 44000:45000
R> subset(mimic_demo$admissions, .data$subject_id %in% .env$subject_id,
+        subject_id:dischtime)
\end{CodeInput}
\begin{CodeOutput}
   subject_id hadm_id           admittime           dischtime
1:      44083  125157 2112-05-04 08:00:00 2112-05-11 14:15:00
2:      44083  131048 2112-05-22 15:37:00 2112-05-25 13:30:00
3:      44083  198330 2112-05-28 15:45:00 2112-06-07 16:50:00
4:      44154  174245 2178-05-14 20:29:00 2178-05-15 09:45:00
5:      44212  163189 2123-11-24 14:14:00 2123-12-30 14:31:00
6:      44222  192189 2180-07-19 06:55:00 2180-07-20 13:00:00
7:      44228  103379 2170-12-15 03:14:00 2170-12-24 18:00:00
\end{CodeOutput}
\end{CodeChunk}

By using \pkg{rlang} pronouns (\texttt{.data} and \texttt{.env}), the
distinction can readily be made between a name referring to an object
within the context of the data and an object within the context of the
calling environment.

\hypertarget{data-source-set-up}{%
\subsubsection{Data source set-up}\label{data-source-set-up}}

In order to make a dataset accessible to \pkg{ricu}, three steps are
necessary, each handled by an exported S3 generic function:
\texttt{download\_scr()}, \texttt{import\_src()} and
\texttt{attach\_src()}. The first two steps, download and import, are
one-time procedures, whereas attaching is carried out every time the
package namespace is loaded. By default, all data sources known to
\pkg{ricu} are configured to be attached and in case some data is
missing for a given data source, the missing data is downloaded and
imported on first access. For data download, several environment
variables can be configured:

\begin{itemize}
\tightlist
\item
  \texttt{RICU\_PHYSIONET\_USER}/\texttt{RICU\_PHYSIONET\_PASS}:
  PhysioNet user name and password with access to the requested dataset.
\item
  \texttt{RICU\_AUMC\_TOKEN}: Download token, extracted from the
  download URL received when requesting data access.
\end{itemize}

If any of the required access credentials are not available as
environment variables, the user is queried in interactive sessions. Each
of the datasets requires 5-10 GB disk space for permanent storage.
Additionally, 50-100 GB of temporary disk storage is required during
download and import of any of the datasets. Memory requirements are kept
low by performing all set-up operations only on subsets of rows at the
time, such that 8 GB of memory should suffice. Initial data source set
up (depending on available download speeds and CPU/disk type) may take
upwards of an hour per dataset.

Further environment variables can be set to customize certain aspects of
\pkg{ricu} data handling:

\begin{itemize}
\tightlist
\item
  \texttt{RICU\_DATA\_PATH}: Data storage location (can be queried by
  calling \texttt{data\_dir()}).
\item
  \texttt{RICU\_CONFIG\_PATH}: Comma-separated paths to directories
  containing configuration files (in addition to the default location;
  retrievable using \texttt{config\_paths()}).
\item
  \texttt{RICU\_SRC\_LOAD}: Comma-separated data source names that are
  set up for being automatically attached on namespace loading (the
  current set of data sources is available as
  \texttt{auto\_attach\_srcs()}).
\end{itemize}

After successful data download, importing prepares tables for efficient
random row-access, for which the raw data format (.csv) is not well
suited. Tables are read in using \pkg{readr} \citep{hester2020},
potentially (re-)partitioned row-wise, and re-saved using \pkg{fst}.
Finally, attaching a dataset creates a corresponding \texttt{src\_env}
object which together with associated meta-data is used by \pkg{ricu} to
run queries against the data.

\hypertarget{data-loading}{%
\subsubsection{Data loading}\label{data-loading}}

The lowest level of data access is direct subsetting of
\texttt{src\_tbl} objects as shown at the start of this section
(\ref{implementation-details}). Building on that, several S3 generic
functions successively homogenize data representations, starting with
\texttt{load\_src()}, which provides a string-based interface to
\texttt{subset()} for all but the row-subsetting expression.

\begin{CodeChunk}
\begin{CodeInput}
R> load_src("admissions", "mimic_demo", subject_id > 44000,
+          cols = c("hadm_id", "admittime", "dischtime"))
\end{CodeInput}
\begin{CodeOutput}
   hadm_id           admittime           dischtime
1:  125157 2112-05-04 08:00:00 2112-05-11 14:15:00
2:  131048 2112-05-22 15:37:00 2112-05-25 13:30:00
3:  198330 2112-05-28 15:45:00 2112-06-07 16:50:00
4:  174245 2178-05-14 20:29:00 2178-05-15 09:45:00
5:  163189 2123-11-24 14:14:00 2123-12-30 14:31:00
6:  192189 2180-07-19 06:55:00 2180-07-20 13:00:00
7:  103379 2170-12-15 03:14:00 2170-12-24 18:00:00
\end{CodeOutput}
\end{CodeChunk}

As data sources differ in their representation of time-stamps, a next
step in data homogenization is to converge to a common format: the time
difference to the origin time-point of a given ID system.

\begin{CodeChunk}
\begin{CodeInput}
R> load_difftime("admissions", "mimic_demo", subject_id > 44000,
+               cols = c("hadm_id", "admittime", "dischtime"))
\end{CodeInput}
\end{CodeChunk}

\begin{CodeChunk}
\begin{CodeOutput}
# An `id_tbl`: 7 x 3
# Id var:      `hadm_id`
  hadm_id admittime dischtime
    <int> <drtn>    <drtn>
1  103379 0 mins    13846 mins
2  125157 0 mins    10455 mins
3  131048 0 mins     4193 mins
4  163189 0 mins    51857 mins
5  174245 0 mins      796 mins
6  192189 0 mins     1805 mins
7  198330 0 mins    14465 mins
\end{CodeOutput}
\end{CodeChunk}

The function \texttt{load\_difftime()} is expected to return timestamps
as base R \texttt{difftime} vectors (using \texttt{mins} as time unit).
The argument \texttt{id\_hint} can be used to specify a preferred ID
system but if not available in raw data, \texttt{load\_difftime()} will
return data using the ID system with highest cardinality. In the above
example, if \texttt{icustay\_id} were requested, data would be returned
using \texttt{hadm\_id}, whereas the a \texttt{subject\_id} request
would be honored, as these two ID columns are available for the
\texttt{admissions} table.

Building on \texttt{load\_difftime()} functionality, \texttt{load\_id()}
(and analogously \texttt{load\_ts()}) returns an \texttt{id\_tbl} (or
\texttt{ts\_tbl}) object with the requested ID system (passed as
\texttt{id\_var} argument). This uses raw data IDs if available or calls
\texttt{change\_id()} in order to convert to the desired ID system.
Similarly, where \texttt{load\_difftime()} returns data with fixed time
interval of one minute, \texttt{load\_id()} allows for arbitrary time
intervals (using \texttt{change\_interval()}).

\begin{CodeChunk}
\begin{CodeInput}
R> load_id("admissions", "mimic_demo", subject_id > 44000,
+         cols = c("admittime", "dischtime"), id_var = "hadm_id")
\end{CodeInput}
\end{CodeChunk}

\begin{CodeChunk}
\begin{CodeOutput}
# An `id_tbl`: 7 x 3
# Id var:      `hadm_id`
  hadm_id admittime dischtime
    <int> <drtn>    <drtn>
1  103379 0 hours   230 hours
2  125157 0 hours   174 hours
3  131048 0 hours    69 hours
4  163189 0 hours   864 hours
5  174245 0 hours    13 hours
6  192189 0 hours    30 hours
7  198330 0 hours   241 hours
\end{CodeOutput}
\end{CodeChunk}

\hypertarget{data-source-configuration}{%
\subsubsection{Data source
configuration}\label{data-source-configuration}}

Data source environments (and corresponding \texttt{src\_tbl} objects)
are constructed using source configuration objects: list-based
structures, inheriting from \texttt{src\_cfg} and from any number of
data source-specific class names with suffix \texttt{\_cfg} appended (as
discussed at the beginning of Section \ref{implementation-details}). The
exported function \texttt{load\_src\_cfg()} reads a JSON formatted file
using \pkg{jsonlite} \citep{ooms2014}, and creates a \texttt{src\_cfg}
object per datasource and further therein contained objects.

\begin{CodeChunk}
\begin{CodeInput}
R> cfg <- load_src_cfg("mimic_demo")
R> str(cfg, max.level = 2L, width = 70L)
\end{CodeInput}
\begin{CodeOutput}
List of 1
 $ mimic_demo:List of 6
  ..$ name   : chr "mimic_demo"
  ..$ prefix : chr [1:2] "mimic_demo" "mimic"
  ..$ id_cfg : id_cfg [1:3] `subject_id`, `hadm_id`, `icustay_id`
  ..$ col_cfg: col_cfg [1:25] [0, 0, 5, 0, 1], [0, 1, 6, 0, 1], [1, 0, 0...
  ..$ tbl_cfg: tbl_cfg [1:25] [?? x 19; 1], [?? x 24; 1], [?? x 4; 1], [...
  ..$ extra  :List of 1
  ..- attr(*, "class")= chr [1:3] "mimic_demo_cfg" "mimic_cfg" "src_cfg"
\end{CodeOutput}
\begin{CodeInput}
R> mi_cfg <- cfg[["mimic_demo"]]
\end{CodeInput}
\end{CodeChunk}

In addition to required fields \texttt{name} and \texttt{prefix} (used
as class prefix), as well as further arbitrary fields (\texttt{url} in
this case), several additional configuration objects are part of
\texttt{src\_cfg}: \texttt{id\_cfg}, \texttt{col\_cfg} and
\texttt{tbl\_cfg}.

\hypertarget{id-configuration}{%
\paragraph{ID configuration}\label{id-configuration}}

An \texttt{id\_cfg} object contains an ordered set of key-value pairs
representing patient ID systems in a dataset. An implicit assumption
currently is that a given patient ID system is used consistently
throughout a dataset, meaning that for example an ICU stay ID is always
referred to by the same name throughout all tables containing a
corresponding column. Owing to the relational origins of these datasets
this has been fulfilled in all instances encountered so far. In
MIMIC-III, ID systems

\begin{CodeChunk}
\begin{CodeInput}
R> as_id_cfg(mi_cfg)
\end{CodeInput}
\begin{CodeOutput}
<id_cfg<mimic_demo[patient < hadm < icustay]>[3]>
     patient         hadm      icustay 
`subject_id`    `hadm_id` `icustay_id` 
\end{CodeOutput}
\end{CodeChunk}

are available, allowing for identification of individual patients, their
(potentially multiple) hospital admissions over the course of the years
and their corresponding ICU admissions (as well as potential
re-admissions). Ordering corresponds to cardinality: moving to larger
values implies moving along a one-to-many relationship. This information
is used in data-loading, whenever the target ID system is not contained
in the raw data.

\hypertarget{default-column-configuration}{%
\paragraph{Default column
configuration}\label{default-column-configuration}}

Again used in data loading, this per-table set of key-value pairs
specifies column defaults as \texttt{col\_cfg} object. Each key
describes a type of column with special meaning and the corresponding
value specifies said column for a given table.

\begin{CodeChunk}
\begin{CodeInput}
R> as_col_cfg(mi_cfg)
\end{CodeInput}
\begin{CodeOutput}
<col_cfg<mimic_demo[id_var, index_var, time_vars, unit_var, val_var]>[25]>
        admissions            callout         caregivers        chartevents 
   [0, 0, 5, 0, 1]    [0, 1, 6, 0, 1]    [1, 0, 0, 0, 1]    [0, 1, 2, 1, 1] 
         cptevents              d_cpt    d_icd_diagnoses   d_icd_procedures 
   [0, 1, 1, 0, 1]    [1, 0, 0, 0, 1]    [1, 0, 0, 0, 1]    [1, 0, 0, 0, 1] 
           d_items         d_labitems     datetimeevents      diagnoses_icd 
   [1, 0, 0, 0, 1]    [1, 0, 0, 0, 1]    [0, 1, 3, 0, 1]    [0, 0, 0, 0, 1] 
          drgcodes           icustays     inputevents_cv     inputevents_mv 
   [0, 0, 0, 0, 1]    [0, 1, 2, 0, 1]    [0, 1, 2, 1, 1]    [0, 1, 4, 1, 1] 
         labevents microbiologyevents       outputevents           patients 
   [0, 1, 1, 1, 1]    [0, 1, 2, 0, 1]    [0, 1, 2, 1, 1]    [0, 0, 4, 0, 1] 
     prescriptions procedureevents_mv     procedures_icd           services 
   [0, 1, 2, 1, 1]    [0, 1, 4, 1, 1]    [0, 0, 0, 0, 1]    [0, 1, 1, 0, 1] 
         transfers 
   [0, 1, 2, 0, 1] 
\end{CodeOutput}
\end{CodeChunk}

The following column defaults are currently in use throughout \pkg{ricu}
but the set of keys can be extended to arbitrary new values:

\begin{itemize}
\tightlist
\item
  \texttt{id\_var}: In case a table does not contain at least one ID
  column corresponding to one of the ID systems specified as
  \texttt{id\_cfg}, the default ID column can be set on a per-table
  basis as \texttt{id\_var}\footnote{This for example is the case for
    the \texttt{d\_items} table in MIMIC-III, which does not contain any
    patient related data, but holds information on items encoding types
    of measurements, procedures, etc., used throughout other tables
    holding actual patient data for identifying the type data point, in
    line with the relational structure of the data source.}.
\item
  \texttt{index\_var}: A column that is used to define an ordering in
  time over rows, thereby providing a time-series index\footnote{For the
    MIMIC-III table \texttt{inputevents\_mv}, one of the four available
    time variables lends itself to be used as index variable more than
    the other candidates and therefore is set as default.}.
\item
  \texttt{time\_vars}: Columns which will be treated as time variables
  (important for converting between ID systems for example), but not as
  time-series indices\footnote{In case of the \texttt{admissions} table
    in MIMIC-III for example, a total of five columns are considered to
    be time variables, none of which stands out as potential
    \texttt{index\_var}.}.
\item
  \texttt{unit\_var}: Used in concept loading (more specifically for
  \texttt{num\_cncpt} concepts, see Section \ref{concept-specification})
  to identify columns that represent unit of measurement information.
\item
  \texttt{val\_var}: Again used when loading data concepts, this
  identified a default value variable in a table, representing the
  column of interest to be used as returned data column.
\end{itemize}

While \texttt{id\_var}, \texttt{index\_var} and \texttt{time\_vars} are
used to provide sensible defaults to functions used for general data
loading (Section \ref{data-loading}), \texttt{unit\_var},
\texttt{val\_var}, as well as potential user-defined defaults are only
used in concept loading (see Section \ref{clinical-concepts}) and
therefore need not be prioritized when integrating new data sources
until data concepts have been mapped.

\hypertarget{table-configuration}{%
\paragraph{Table configuration}\label{table-configuration}}

Finally, \texttt{tbl\_cfg} objects are used during the initial set-up of
a data source. In order to create a representation of a table that is
accessible from \pkg{ricu} from raw data, several key pieces of
information are required:

\begin{itemize}
\item
  File name(s): In the simplest case, a single file corresponds to a
  single table. Other scenarios that have been encountered (and are
  therefore handled) include tables partitioned into multiple files and
  .tar archives containing multiple tables.
\item
  Column specification: For each column, the expected data type has to
  be known, as well as a pair of names, one corresponding to the raw
  data column name and one corresponding to the column name to be used
  within \pkg{ricu}.
\item
  (Optional) number of rows: Used as sanity check whenever available.
\item
  (Optional) partitioning information: For very \emph{long} tables it
  can be useful to specify a row-partitioning. This currently is only
  possible by applying a vector of breakpoints to a single numeric
  column, thereby defining a grouping.
\end{itemize}

\begin{CodeChunk}
\begin{CodeInput}
R> as_tbl_cfg(mi_cfg)
\end{CodeInput}
\begin{CodeOutput}
<tbl_cfg<mimic_demo[rows x cols; partitions]>[25]>
        admissions            callout         caregivers        chartevents 
      [?? x 19; 1]       [?? x 24; 1]        [?? x 4; 1]       [?? x 15; 2] 
         cptevents              d_cpt    d_icd_diagnoses   d_icd_procedures 
      [?? x 12; 1]        [?? x 9; 1]        [?? x 4; 1]        [?? x 4; 1] 
           d_items         d_labitems     datetimeevents      diagnoses_icd 
      [?? x 10; 1]        [?? x 6; 1]       [?? x 14; 1]        [?? x 5; 1] 
          drgcodes           icustays     inputevents_cv     inputevents_mv 
       [?? x 8; 1]       [?? x 12; 1]       [?? x 22; 1]       [?? x 31; 1] 
         labevents microbiologyevents       outputevents           patients 
       [?? x 9; 1]       [?? x 16; 1]       [?? x 13; 1]        [?? x 8; 1] 
     prescriptions procedureevents_mv     procedures_icd           services 
      [?? x 19; 1]       [?? x 25; 1]        [?? x 5; 1]        [?? x 6; 1] 
         transfers 
      [?? x 13; 1] 
\end{CodeOutput}
\end{CodeChunk}

For the \texttt{chartevents} table of the MIMIC-III demo dataset, for
example, rows are partitioned into two groups, while all other tables
are represented by a single partition. Furthermore, the expected number
of rows is unknown (\texttt{??}) as this is missing from the
corresponding \texttt{tbl\_cfg} object.

\hypertarget{adding-external-datasets}{%
\subsection{Adding external datasets}\label{adding-external-datasets}}

In order to add a new dataset to \pkg{ricu}, several aspects outlined in
the previous subsections require consideration. For illustration
purposes, code for integrating AmsterdamUMCdb as external dataset is
available from \href{https://github.com/eth-mds/aumc}{Github}. While
this is no longer needed for using the \texttt{aumc} data source, the
repository will remain as it might serve as template to integration of
new datasets.

Using a configuration file as described in Section
\ref{data-source-configuration} and pointing the
\texttt{RICU\_CONFIG\_PATH} environment variable at its location, data
can be prepared for use with \pkg{ricu} using \texttt{import\_src()} and
made available to \pkg{ricu} using \texttt{attach\_src()}. In addition
to providing a configuration file, dataset-specific implementations of
some of the S3 generic functions involved in data-loading might be
required. In the case of AmsterdamUMCdb, a class-specific implementation
of \texttt{load\_difftime()} is required, as raw time-stamps are
recorded in milliseconds (instead of minutes), as well as a
class-specific implementation of the S3 generic function
\texttt{id\_win\_helper()}\footnote{This method is used as part of
  \texttt{change\_id()} which is called whenever the requested ID system
  is not available in raw data. A component central to conversion
  between patient ID systems is a table which contains patient IDs as
  columns alongside columns with start and end-points for each, thereby
  specifying a mapping between ID systems. Dataset-specific construction
  of such a table is handled by the S3 generic function
  \texttt{id\_win\_helper()}. As construction of such tables can be
  expensive (involving several merge operations of tables with
  10\textsuperscript{4}-10\textsuperscript{5} rows) and used frequently
  (potentially with every single data request), the resulting table is
  cached in memory with session persistence.}.

\hypertarget{data-concepts}{%
\section{Data concepts}\label{data-concepts}}

One of the key components of \pkg{ricu} is a scheme for specifying how
to retrieve data corresponding to pre-defined clinical concepts from a
given data source, in turn enabling dataset agnostic code for analysis.
Heart rate, for example can be loaded using the \texttt{hr} concept as

\begin{CodeChunk}
\begin{CodeInput}
R> load_concepts("hr", c("mimic_demo", "eicu_demo"), verbose = FALSE)
\end{CodeInput}
\begin{CodeOutput}
# A `ts_tbl`: 152,197 x 4
# Id vars:    `source`, `icustay_id`
# Units:      `hr` [bpm]
# Index var:  `charttime` (1 hours)
        source     icustay_id charttime    hr
        <chr>           <int> <drtn>    <dbl>
      1 eicu_demo      141764   0 hours   119
      2 eicu_demo      141764   1 hours   105
      3 eicu_demo      141764   2 hours    95
      4 eicu_demo      141764   3 hours   106
      5 eicu_demo      141764   4 hours   102
    ...
152,193 mimic_demo     298685 314 hours    60
152,194 mimic_demo     298685 315 hours    56
152,195 mimic_demo     298685 316 hours    50
152,196 mimic_demo     298685 317 hours    48
152,197 mimic_demo     298685 318 hours     0
# ... with 152,187 more rows
\end{CodeOutput}
\end{CodeChunk}

This requires some form of infrastructure for concisely specifying how
to retrieve data subsets (Section \ref{concept-specification}), which is
both extensible (to new concepts and new datasets) and flexible enough
to handle concept-specific pre-processing. Additionally, \pkg{ricu} has
included a dictionary with over 100 concepts implemented for all four
supported datasets (where possible; see also Section
\ref{clinical-concepts}). A quick remark on terminology before diving
into more details on how to specify data concepts: A \emph{concept}
corresponds to a clinical variable such as a bilirubin measurement or
the ventilation status of a patient, and an \emph{item} encodes how to
retrieve data corresponding to a given concept from a data source. A
\emph{concept} therefore contains several \emph{items} (zero, one or
several are possible per data source).

\hypertarget{concept-specification}{%
\subsection{Concept specification}\label{concept-specification}}

Similarly to data source configuration (discussed in Section
\ref{data-source-configuration}), concept specification relies on
JSON-formatted text files. A default dictionary of concepts is included
with \pkg{ricu} containing a selection of commonly used clinical
concepts. Several types of concepts exist within \pkg{ricu} and with
extensibility in mind, new types can easily be added.

All concepts consist of minimal meta-data including a name, target class
(defaults to \texttt{ts\_tbl}; see Section \ref{data-classes}), an
aggregation specification\footnote{Every concept needs a default
  aggregation method which can be used during data loading to return
  data that is unique per key (either per \texttt{id\_vars} group or per
  combination of \texttt{ìd\_vars} and \texttt{index\_var}) otherwise
  down-stream merging of multiple concepts is ill-defined. The
  aggregation default can be overridden during loading or as
  specification of a \texttt{rec\_cncpt} object. If no aggregation
  method is explicitly indicated the global default is \texttt{first()}
  for character, \texttt{median()} for numeric and \texttt{sum()} for
  logical vectors. For logical data, if a concept of type
  \texttt{lgl\_cncpt} is used, the count of \texttt{TRUE} values is
  converted back to logical, thereby providing \texttt{any()} type
  functionality.} and class information (defaults to
\texttt{num\_concept}), as well as optional \texttt{description} and
\texttt{category} information. Adding to that, depending on concept
class, further fields can be added. In the case of the most widespread
concept type (\texttt{num\_cncpt}; used to represent numeric data) this
is \texttt{unit} which encodes one (or several synonymous) unit(s) of
measurement, as well as a minimal and maximal plausible values
(specified as \texttt{min} and \texttt{max}). The concept for heart rate
data (\texttt{hr}) for example can be specified as

\begin{verbatim}
{
  "hr": {
    "unit": ["bpm", "/min"],
    "min": 0,
    "max": 300,
    "description": "heart rate",
    "category": "routine vital signs",
    "sources": {
      ...
    }
  }
}
\end{verbatim}

Meta-data is used during concept loading for data-preprocessing. For
numeric concepts, the specified measurement unit is compared to that of
the data (if available), with messages being displayed in case of
mismatches, while the range of plausible values is used to filter out
measurements that fall outside the specified interval. Other types of
concepts include categorical concepts (\texttt{fct\_cncpt}), concept
representing binary data (\texttt{lgl\_cncpt}), as well as recursive
concepts (\texttt{rec\_cncpt}), which build on other \emph{atomic}
concepts\footnote{An example for a recursive concept is the
  PaO\textsubscript{2}/FiO\textsubscript{2} ratio, used for instance to
  assess patients with acute respiratory distress syndrome (ARDS) or for
  sepsis-related organ failure assessment (SOFA)
  \citep{villar2013, vincent1996}. Given both PaO\textsubscript{2} and
  FiO\textsubscript{2} as individual concepts, the
  PaO\textsubscript{2}/FiO\textsubscript{2} ratio is provided by
  \pkg{ricu} as a recursive concept (\texttt{pafi}), requesting the two
  atomic concepts \texttt{pao2} and \texttt{fio2} and performing some
  form of imputation for when at a given time step one or both values
  are missing.}.

Specification of how data can be retrieved from a data source is encoded
by data \emph{items}. Lists of data items (associated with data source
names) are provided as \texttt{sources} element (instead of \texttt{...}
in the above code block). For the demo datasets corresponding eICU and
MIMIC-III, heart rate data retrieval is specified as

\begin{verbatim}
{
  "eicu_demo": [
    {
      "table": "vitalperiodic",
      "val_var": "heartrate",
      "class": "col_itm"
    }
  ],
  "mimic_demo": [
    {
      "ids": [211, 220045],
      "table": "chartevents",
      "sub_var": "itemid"
    }
  ]
}
\end{verbatim}

Analogously to how different types of concepts are used to represent
different types of data, different types of items handle different types
of data loading. The most common scenario is selecting a subset of rows
from a table by matching a set of ID values (\texttt{sub\_itm}). In the
above example, heart rate data in MIMIC-III can be located by searching
for ID values 211 and 220045 in column \texttt{itemid} of table
\texttt{chartevents} (heart rate data is stored in \emph{long} format).
Conversely, heart rate data in eICU is stored in \emph{wide} format,
requiring no row-subsetting. Column \texttt{heartrate} of table
\texttt{vitalperiodic} contains all corresponding data and such data
situations are handled by the \texttt{col\_itm} class. Other item
classes include \texttt{rgx\_itm} where a regular expression is used for
selecting rows and \texttt{fun\_itm} where an arbitrary function can be
used for data loading. If a data loading scenario is not covered by
these classes, adding further \texttt{itm} subclasses is encouraged.

In order to extend the current concept library both to new datasets and
new concepts, further JSON files can be incorporated by adding their
paths to \texttt{RICU\_CONFIG\_PATH}. Concepts with names that already
exist are only used for their \texttt{sources} entries, such that
\texttt{hr} for \texttt{new\_dataset} can be specified as

\begin{verbatim}
"hr": {
  "sources": {
    "new_dataset": [
      {
        "ids": 6640,
        "table": "numericitems",
        "sub_var": "itemid"
      }
    ]
  }
}
\end{verbatim}

whereas concepts with non-existing names are treated as new concepts.

Central to providing the required flexibility for loading of certain
data concepts that require some specific pre-processing are callback
functions that can be specified for several \emph{item} types. Functions
(with appropriate signatures), designated as \texttt{callback}
functions, are invoked on individual data items, before concept-related
preprocessing is applied. A common scenario for this is unit of
measurement conversion: In MIMIC-III data for example, several
\texttt{itemid} values correspond to temperature measurements, some of
which refer to temperatures measured in degrees Celsius whereas others
are used for measurements in degrees Fahrenheit. As the information
encoding which measurement corresponds to which \texttt{itemid} values
is no longer available during concept-related preprocessing, this is
best resolved at the level of individual data items. Several function
factories are available for generating callback functions and
\texttt{convert\_unit()} is intended for covering unit conversions. Data
\emph{items} corresponding to the \texttt{temp} concept for MIMIC-III
are specified as

\begin{verbatim}
{
  "mimic_demo": [
    {
      "ids": [676, 677, 223762],
      "table": "chartevents",
      "sub_var": "itemid"
    },
    {
      "ids": [678, 679, 223761, 224027],
      "table": "chartevents",
      "sub_var": "itemid",
      "callback": "convert_unit(fahr_to_cels, 'C', 'f')"
    }
  ]
}
\end{verbatim}

indicating that for ID values 676, 677 and 223762 no pre-processing is
required and for the remaining ID values the function
\texttt{fahr\_to\_cels()} is applied to entries of the \texttt{val\_var}
column where the regular expression \texttt{"f"} is \texttt{TRUE} for
the \texttt{unit\_var} column (the values of which being ultimately
replaced with \texttt{"C"}).

\hypertarget{data-classes}{%
\subsection{Data classes}\label{data-classes}}

In order to represent tabular ICU data, \pkg{ricu} provides several
classes, all inheriting from \texttt{data.table}. The most basic of
which, \texttt{id\_tbl}, marks one (or several) columns as
\texttt{id\_vars} which serve to define a grouping (i.e.~identify
patients or unit stays). Inheriting from \texttt{id\_tbl},
\texttt{ts\_tbl} is capable of representing grouped time-series data. In
addition to \texttt{id\_var} column(s), a single column is marked as
\texttt{index\_var} and is required to hold a base R \texttt{difftime}
vector. Furthermore, \texttt{ts\_tbl} contains a scalar-valued
\texttt{difftime} object as \texttt{interval} attribute, specifying the
time-series step size\footnote{As further extension, a \texttt{win\_tbl}
  object is being considered for inclusion, capable of representing time
  intervals. Such an object could prove convenient for example when
  dealing with infusions as infusion parameters such as medication rate
  frequently are specified with explicit begin and end times (see
  Section \ref{treatment-related-information}).}.

Meta data is transiently added to \texttt{data.table} objects by classes
inheriting from \texttt{id\_tbl} and S3 generic functions which allow
for object modifications, down-casting is implicit:

\begin{CodeChunk}
\begin{CodeInput}
R> (dat <- ts_tbl(a = 1:5, b = hours(1:5), c = rnorm(5)))
\end{CodeInput}
\begin{CodeOutput}
# A `ts_tbl`: 5 x 3
# Id var:     `a`
# Index var:  `b` (1 hours)
      a b            c
  <int> <drtn>   <dbl>
1     1 1 hours  0.315
2     2 2 hours -1.84
3     3 3 hours  1.73
4     4 4 hours -1.01
5     5 5 hours -0.419
\end{CodeOutput}
\begin{CodeInput}
R> dat[["b"]] <- dat[["b"]] + mins(30)
R> dat
\end{CodeInput}
\begin{CodeOutput}
# An `id_tbl`: 5 x 3
# Id var:      `a`
      a b               c
  <int> <drtn>      <dbl>
1     1  5400 secs  0.315
2     2  9000 secs -1.84
3     3 12600 secs  1.73
4     4 16200 secs -1.01
5     5 19800 secs -0.419
\end{CodeOutput}
\end{CodeChunk}

Due to time-series step size of \texttt{dat} being specified as 1 hour,
an internal inconsistency is encountered when shifting time stamps by 30
minutes, as time-steps are no longer multiples of the time-series
interval, in turn causing down-casting to \texttt{id\_tbl}. If column
\texttt{a} were to be removed, direct down-casting to
\texttt{data.table} would be required in order to resolve
inconsistencies\footnote{Updating an object inheriting from
  \texttt{id\_tbl} using \texttt{data.table::set()} bypasses consistency
  checks as this is not an S3 generic function and therefore its
  behavior cannot be tailored to requirements of \texttt{id\_tbl}
  objects. It therefore is up to the user to avoid vitiating
  \texttt{id\_tbl} objects in such a way.}.

Utilizing the attached meta-data, several utility functions can be
called with concise semantics. This includes functions for sorting,
checking for duplicates, aggregating data per combination of
\texttt{id\_vars} (and time-step), checking time series data for gaps,
verifying whether the time-series is regular and converting between
irregular and regular time-series, as well as functions for several
types of moving window operations. Adding to those class-specific
implementations, \texttt{id\_tbl} objects inherit from
\texttt{data.table} (and therefore from \texttt{data.frame}), ensuring
compatibility with a wide range of functionality targeted at these
base-classes.

\hypertarget{clinical-concepts}{%
\subsection{Clinical concepts}\label{clinical-concepts}}

The current selection of clinical concepts that is included with
\pkg{ricu} covers many physiological variables that are available
throughout the included datasets. Treatment-related information on the
other hand, being more heterogeneous in nature and therefore harder to
harmonize across datasets, has been added on an as-needed basis and
therefore is more limited in breadth.

Available concepts can be enumerated using \texttt{load\_dictionary()}
and the utility function \texttt{explain\_dictionary()} can be used to
display some concept meta-data.

\begin{CodeChunk}
\begin{CodeInput}
R> dict <- load_dictionary(c("mimic_demo", "eicu_demo"))
R> head(dict)
\end{CodeInput}
\begin{CodeOutput}
<concept[6]>
                                  abx                              adh_rate 
           antibiotics <lgl_cncpt[4]>       vasopressin rate <num_cncpt[3]> 
                                  adm                                   age 
patient admission type <fct_cncpt[2]>            patient age <num_cncpt[2]> 
                                  alb                                   alp 
               albumin <num_cncpt[2]>   alkaline phosphatase <num_cncpt[2]> 
\end{CodeOutput}
\begin{CodeInput}
R> explain_dictionary(head(dict))
\end{CodeInput}
\begin{CodeOutput}
      name     category            description
1      abx  medications            antibiotics
2 adh_rate  medications       vasopressin rate
3      adm demographics patient admission type
4      age demographics            patient age
5      alb    chemistry                albumin
6      alp    chemistry   alkaline phosphatase
\end{CodeOutput}
\end{CodeChunk}

The following sub-sections serve to introduce some of the included
concepts as well as highlight limitations that come with current
implementations. Grouping the available concepts by category yields the
following counts

\begin{CodeChunk}
\begin{CodeInput}
R> table(vapply(dict, `[[`, character(1L), "category"))
\end{CodeInput}
\begin{CodeOutput}

   blood gas    chemistry demographics   hematology  medications microbiology 
          10           21            6           20           16            1 
neurological      outcome       output  respiratory       vitals 
           8           19            2           10            6 
\end{CodeOutput}
\end{CodeChunk}

\hypertarget{physiological-data}{%
\subsubsection{Physiological data}\label{physiological-data}}

The largest and most well established group of concepts (covering more
than half of all currently included concepts) includes physiological
patient measurements such as routine vital signs, respiratory variables,
fluid discharge amounts, as well as many kinds of laboratory tests
including blood gas measurements, chemical analysis of body fluids and
hematology assays.

\begin{CodeChunk}
\begin{CodeInput}
R> load_concepts(c("alb", "glu"), "mimic_demo", interval = mins(15L),
+               verbose = FALSE)
\end{CodeInput}
\begin{CodeOutput}
# A `ts_tbl`: 1,965 x 4
# Id var:     `icustay_id`
# Units:      `alb` [g/dL], `glu` [mg/dL]
# Index var:  `charttime` (15 mins)
      icustay_id charttime    alb   glu
           <int> <drtn>     <dbl> <dbl>
    1     201006 -3495 mins  NA     116
    2     201006 -2745 mins  NA      83
    3     201006 -1275 mins  NA      91
    4     201006    15 mins   2.4   175
    5     201006   675 mins  NA     129
  ...
1,961     298685 15600 mins  NA     159
1,962     298685 16365 mins   2.2   153
1,963     298685 17400 mins  NA     182
1,964     298685 17595 mins  NA     122
1,965     298685 17955 mins   2.5   121
# ... with 1,955 more rows
\end{CodeOutput}
\end{CodeChunk}

Most concepts of this kind are represented by \texttt{num\_cncpt}
objects with an associated unit of measurement and a range of
permissible values. Data is mainly returned as \texttt{ts\_tbl} objects,
representing time-dependent observations. Apart from conversion to a
common unit (possibly using the \texttt{convert\_unit()} callback
function), little has to be done in terms of pre-processing: values are
simply reported at time-points rounded to the requested interval.

\hypertarget{patient-demographics}{%
\subsubsection{Patient demographics}\label{patient-demographics}}

Moving on from dynamic, time-varying patient data, this group of
concepts focuses on static patient information. While the assumption of
remaining constant throughout a stay is likely to hold for variables
such as patient sex or height this is only approximately true for others
including age or weight. Nevertheless such effects are ignored and
concepts of this group will be mainly returned as \texttt{id\_tbl}
objects with no corresponding time-stamps included.

Whenever requesting concepts which are returned with associated
time-stamps (e.g.~glucose) alongside time-constant data (e.g.~age),
merging will duplicate static data over all time-points.

\begin{CodeChunk}
\begin{CodeInput}
R> load_concepts(c("age", "glu"), "mimic_demo", verbose = FALSE)
\end{CodeInput}
\begin{CodeOutput}
# A `ts_tbl`: 1,914 x 4
# Id var:     `icustay_id`
# Units:      `age` [years], `glu` [mg/dL]
# Index var:  `charttime` (1 hours)
      icustay_id charttime   age   glu
           <int> <drtn>    <dbl> <dbl>
    1     201006 -58 hours  68.9   116
    2     201006 -45 hours  68.9    83
    3     201006 -21 hours  68.9    91
    4     201006   0 hours  68.9   175
    5     201006  11 hours  68.9   129
  ...
1,910     298685 260 hours  80.1   159
1,911     298685 272 hours  80.1   153
1,912     298685 290 hours  80.1   182
1,913     298685 293 hours  80.1   122
1,914     298685 299 hours  80.1   121
# ... with 1,904 more rows
\end{CodeOutput}
\end{CodeChunk}

Despite a best-effort approach, data availability can be a limiting
factor. While for physiological variables, there is good agreement even
across continents, data-privacy considerations, as well as lack of a
common standard for data encoding, may cause issues that are hard to
resolve. In some cases, this can be somewhat mitigated while in others,
this is a limitation to be kept in mind. In AmsterdamUMCdb, for example,
patient age, height and weight are not available as continuous
variables, but as factor with patients binned into groups. Such
variables are then approximated by returning the respective mid-points
of groups for \texttt{aumc} data\footnote{Prioritizing consistency over
  accuracy, one could apply the same binning to datasets which report
  numeric values, but the concepts included with \pkg{ricu} attempt to
  strike a balance between consistency and amount of applied
  pre-processing. With the extensible architecture of data concepts,
  however, such categorical variants of patient demographic concepts
  could easily be added.}. Other concepts, such as \texttt{adm}
(categorizing admission types) or a prospective \texttt{icd} concept
(diagnoses as ICD-9 codes) can only return data if available from the
data source in question. Unfortunately, neither \texttt{aumc} nor
\texttt{hirid} contain ICD-9 encoded diagnoses, and in the case of
\texttt{hirid}, no diagnosis information is available at all.

\hypertarget{treatment-related-information}{%
\subsubsection{Treatment-related
information}\label{treatment-related-information}}

The largest group of concepts dealing with treatment-related information
is described by the \texttt{medications} category. In addition to drug
administrations, only basic ventilation information is currently
provided as ready to use concept. Just like availability of common ICU
procedures, patient medication is also underdeveloped, covering mainly
vasopressor administrations, as well as corticosteroids and antibiotics.
The current concepts retrieving treatment-related information are mostly
focused on providing data required for constructing clinical scores
described in Section \ref{outcomes}.

Ventilation is represented by several concepts: a ventilation indicator
variable (\texttt{vent\_ind}), as well as ventilation durations
(\texttt{vent\_dur}) are constructed from start and events
(\texttt{vent\_start} and \texttt{vent\_end}). This includes any kind of
mechanical ventilation (invasive via an endotracheal or tracheostomy
tube), as well as non-invasive ventilation via face or nasal masks. In
line with other concepts belonging to this group, the current state is
far from being comprehensive and expansion to further ventilation
parameters is desirable.

The singular concept addressing antibiotics (\texttt{abx}) returns an
indicator signaling whenever an antibiotic was administered. This
includes any route of administration (intravenous, oral, topical, etc.)
and does neither report dosage, nor active ingredient. Finally,
vasopressor administration is reported by several concepts representing
different vasoactive drugs (including dopamine, dobutamine, epinephrine,
noreponephrine and vasopressin), as well as different administration
aspects such as rate, duration (and for use in SOFA scoring, rate
administered for at least 60 minutes).

\begin{CodeChunk}
\begin{CodeInput}
R> load_concepts(c("abx", "vent_ind", "norepi_rate", "norepi_dur"),
+               "mimic_demo", verbose = FALSE)
\end{CodeInput}
\begin{CodeOutput}
# A `ts_tbl`: 12,547 x 6
# Id var:     `icustay_id`
# Units:      `norepi_rate` [mcg/kg/min]
# Index var:  `startdate` (1 hours)
       icustay_id startdate abx   vent_ind norepi_rate norepi_dur
            <int> <drtn>    <lgl> <lgl>          <dbl> <drtn>
     1     201006 -40 hours TRUE  NA           NA      NA hours
     2     201006 -16 hours TRUE  NA           NA      NA hours
     3     201006   7 hours TRUE  NA           NA      NA hours
     4     201006   8 hours NA    TRUE         NA      60 hours
     5     201006   9 hours NA    TRUE          0.0460 NA hours
   ...
12,543     298685 612 hours NA    TRUE         NA      NA hours
12,544     298685 613 hours NA    TRUE         NA      NA hours
12,545     298685 614 hours NA    TRUE         NA      NA hours
12,546     298685 615 hours NA    TRUE         NA      NA hours
12,547     298685 616 hours NA    TRUE         NA      NA hours
# ... with 12,537 more rows
\end{CodeOutput}
\end{CodeChunk}

As cautioned in Section \ref{patient-demographics}, variability in data
reporting across datasets can lead to issues: the \texttt{prescriptions}
table included with MIMIC-III, for example, reports time-stamps as dates
only, yielding a discrepancy of up to 24 hours when merged with data
where time-accuracy is on the order of minutes. This effect is somewhat
mitigated by shifting time-stamps from midnight to mid-day, but the
underlying accuracy issue of course remains. Another problem exists with
concepts that attempt to report administration windows, as some datasets
do not describe infusions with clear cut start/endpoints but rather
report infusion parameters at (somewhat) regular time intervals. This
can cause artifacts when the requested time step-size deviates from the
dataset inherent time grid.

\hypertarget{outcomes}{%
\subsubsection{Outcomes}\label{outcomes}}

A group of more loosely associated concepts can be used to describe
patient state. This includes common clinical endpoints, such as death or
length of ICU stay, as well as scoring systems such as SOFA, the
systemic inflammatory response syndrome \citep[SIRS;][]{bone1992}
criterion, the National Early Warning Score \citep[NEWS;][]{jones2012}
and the Modified Early Warning Score \citep[MEWS;][]{subbe2001}.

While the more straightforward outcomes can be retrieved directly from
data, clinical scores often incorporate multiple variables, based upon
which a numeric score is constructed. This can typically be achieved by
using concepts of type \texttt{rec\_cncpt}, specifying the needed
components and supplying a callback function that applies rules for
score construction.

\begin{CodeChunk}
\begin{CodeInput}
R> load_concepts(c("sirs", "death"), "mimic_demo", verbose = FALSE,
+               keep_components = TRUE)
\end{CodeInput}
\begin{CodeOutput}
# A `ts_tbl`: 14,295 x 8
# Id var:     `icustay_id`
# Index var:  `charttime` (1 hours)
       icustay_id charttime  sirs death temp_comp hr_comp resp_comp wbc_comp
            <int> <drtn>    <dbl> <lgl>     <int>   <int>     <int>    <int>
     1     201006 -58 hours     1 NA           NA      NA        NA        1
     2     201006 -45 hours     1 NA           NA      NA        NA        1
     3     201006 -21 hours     1 NA           NA      NA        NA        1
     4     201006 -10 hours     2 NA           NA      NA         1        1
     5     201006   0 hours     3 NA            0       1         1        1
   ...
14,291     298685 314 hours     2 NA            0       0         1        1
14,292     298685 315 hours     2 NA            0       0         1        1
14,293     298685 316 hours     2 NA            0       0         1        1
14,294     298685 317 hours     1 NA            0       0         0        1
14,295     298685 318 hours     1 TRUE          0       0        NA        1
# ... with 14,285 more rows
\end{CodeOutput}
\end{CodeChunk}

Callback functions can become rather involved (especially for more
complex concepts such as SOFA) and may include arbitrary arguments to
tune their behavior. As callback functions to \texttt{rec\_cncpt}
objects are typically called internally from \texttt{load\_concepts()},
arguments not used by \texttt{load\_concepts()}, such as
\texttt{keep\_components} in the above example (causing not only the
score column, but also individual score components to be retained) are
forwarded\footnote{Some care has to be taken as when requesting multiple
  concepts within the same call to \texttt{load\_concepts()} all
  involved callback functions will be called with the same forwarded
  arguments. When for example requesting multiple scores, it is
  currently not possible to enable \texttt{keep\_components} for only a
  subset thereof. Furthermore this set-up implies that}.

\hypertarget{examples}{%
\section{Examples}\label{examples}}

In order to briefly illustrate how \pkg{ricu} could be applied to
real-world clinical questions, two toy examples are provided in the
following sections. While the first example fully relies on data
concepts that are included with \pkg{ricu}, the second one explores both
how some data pre-processing can be added to an existing concept by
creating a new \texttt{rec\_cncpt} and how to create an new data concept
altogether.

\hypertarget{lactate-and-mortality}{%
\subsection{Lactate and mortality}\label{lactate-and-mortality}}

First, we investigate the association of lactate levels and mortality.
This problem has been studied before and it is widely accepted that both
static and dynamic lactate indices are associated with increased
mortality \citep{haas2016, nichol2011, van2013}. In order to model this
relationship, we fit a time-varying proportional hazards Cox model
\citep{therneau2000, therneau2015}, which includes the SOFA score as a
general predictor of illness severity, using MIMIC-III data.
Furthermore, for the sake of this example, we are only interested in
patients admitted from 2008 onwards of ages 25 to 65 years old.

\begin{CodeChunk}
\begin{CodeInput}
R> src <- "mimic"
R> 
R> cohort <- load_id("icustays", src, dbsource == "metavision",
+                   cols = NULL)
R> cohort <- load_concepts("age", src, patient_ids = cohort,
+                         verbose = FALSE)
R> 
R> dat <- load_concepts(c("lact", "death", "sofa", "sex"), src,
+                      patient_ids = cohort[age > 25 & age < 65],
+                      verbose = FALSE)
\end{CodeInput}
\end{CodeChunk}

\begin{CodeChunk}
\begin{CodeInput}
R> dat <- dat[, head(.SD, n = match(TRUE, death, .N)), by = c(id_vars(dat))]
R> dat <- fill_gaps(dat)
R> 
R> dat <- replace_na(dat, c(NA, FALSE), type = c("locf", "const"),
+                   by_ref = TRUE, vars = c("lact", "death"),
+                   by = id_vars(dat))
R> 
R> cox_mod <- coxph(
+   Surv(charttime - 1L, charttime, death) ~ lact + sofa,
+   data = dat
+ )
\end{CodeInput}
\end{CodeChunk}

After loading the data, some minor pre-processing is still required
before modeling: first, we want to make sure we only use data up to (and
including) the hour in which the \texttt{death} flag switches to
\texttt{TRUE}. After that we impute missing values for \texttt{lact}
using a last observation carry forward (locf) scheme (observing the
patient grouping) and we simply replace missing \texttt{death} values
with the value \texttt{FALSE}. The resulting model fit can be visualized
as:

\begin{CodeChunk}


\begin{center}\includegraphics{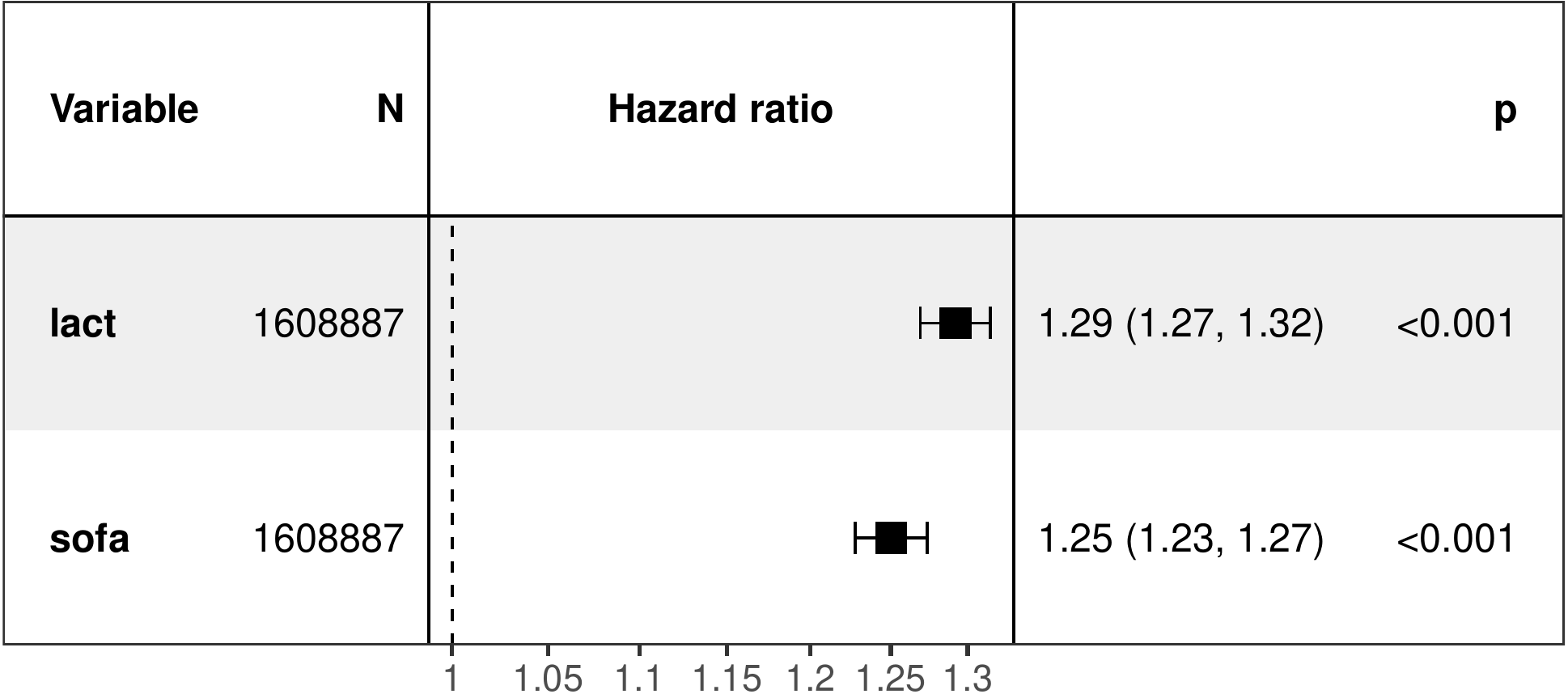} \end{center}

\end{CodeChunk}

A simple exploration already shows that the increased values of lactate
are associated with mortality, even after adjusting for the SOFA score.

\hypertarget{diabetes-and-insulin-treatment}{%
\subsection{Diabetes and insulin
treatment}\label{diabetes-and-insulin-treatment}}

For the next example, again using MIMIC-III data, we turn to the usage
of co-morbidities and treatment related information. We look at the
amount of insulin administered to patients in the first 24 hours from
their ICU admission. In particular, we investigate if diabetic patients
receive more insulin in the first day of their stay compared to
non-diabetic patients. For this we create two concepts: \texttt{ins24},
a binned variable representing the cumulative amount of insulin
administered within the first 24 hours of an ICU admission, and
\texttt{diab}, a logical variable encoding diabetes co-morbidity.

As there already is an insulin concept available, \texttt{ins24} can be
implemented as \texttt{rec\_cncpt}, loading \texttt{ins} with
aggregation set to \texttt{sum()} (instead of \texttt{median()}) and
inserting the callback function \texttt{ins\_cb()} into the loading
process. The callback function takes care of the pre-processing steps
outlined above: first data is subsetted to fall into the the first 24
hours of ICU admissions, followed by binning of summed values.

\begin{CodeChunk}
\begin{CodeInput}
R> ins_breaks <- c(0, 1, 10, 20, 40, Inf)
R> 
R> ins_cb <- function(ins, ...) {
+ 
+   day_one <- function(x) x >= hours(0L) & x <= hours(24L)
+ 
+   idx_var <- index_var(ins)
+   ids_var <- id_vars(ins)
+ 
+   ins <- ins[
+     day_one(get(idx_var)), list(ins24 = sum(ins)), by = c(ids_var)
+   ]
+ 
+   ins <- ins[,
+     ins24 := list(cut(ins24, breaks = ins_breaks, right = FALSE))
+   ]
+ 
+   ins
+ }
R> 
R> ins24 <- load_dictionary(src, "ins")
R> ins24 <- concept("ins24", ins24, "insulin in first 24h", aggregate = "sum",
+                  callback = ins_cb, target = "id_tbl", class = "rec_cncpt")
\end{CodeInput}
\end{CodeChunk}

The binary diabetes concept can be implemented as \texttt{lgl\_cncpt},
for which ICD-9 codes are matched using a regular expression. As we're
not only interested in retrieving diabetic patients, a \texttt{col\_itm}
is more suited for data retrieval over an \texttt{rgx\_itm} and for
creating the required callback function that produces a logical vector
we can use \texttt{transform\_fun()} coupled with a function like
\texttt{grep\_diab()}. The two concepts are then combined using
\texttt{c()} and loaded via \texttt{load\_concepts()}.

\begin{CodeChunk}
\begin{CodeInput}
R> grep_diab <- function(x) grepl("^250\\.?[0-9]{2}$", x)
R> 
R> diab  <- item(src, table = "diagnoses_icd",
+               callback = transform_fun(grep_diab), class = "col_itm")
R> diab  <- concept("diab", diab, "diabetes", target = "id_tbl",
+                  class = "lgl_cncpt")
R> 
R> dat <- load_concepts(c(ins24, diab), id_type = "icustay", verbose = FALSE)
R> dat <- replace_na(dat, "[0,1)", vars = "ins24")
R> dat
\end{CodeInput}
\end{CodeChunk}

After this, we can visualize the difference between the two groups with
a histogram:

\begin{CodeChunk}


\begin{center}\includegraphics{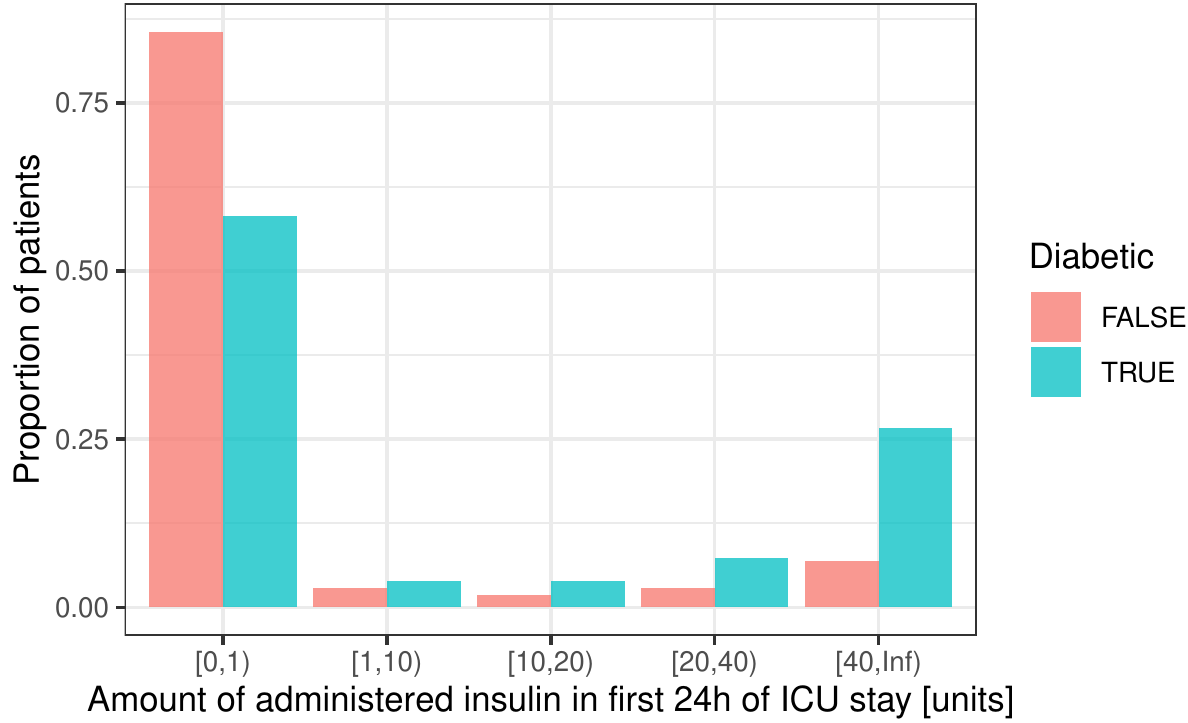} \end{center}

\end{CodeChunk}

The plot suggests that during the first day of ICU stay, perhaps
unsurprisingly, diabetic patients receive more insulin compared to
non-diabetic patients, especially as insulin dose increases.

\hypertarget{acknowledgments}{%
\section{Acknowledgments}\label{acknowledgments}}

Nicolas Bennett, Drago Plečko, Nicolai Meinshausen and Peter Bühlmann
were supported by grant \#2017-110 of the Strategic Focal Area
``Personalized Health and Related Technologies (PHRT)'' of the ETH
Domain for the SPHN/PHRT Driver Project ``Personalized Swiss Sepsis
Study''.


\begin{thebibliography}{35}
\newcommand{\enquote}[1]{``#1''}
\providecommand{\natexlab}[1]{#1}
\providecommand{\url}[1]{\texttt{#1}}
\providecommand{\urlprefix}{URL }
\expandafter\ifx\csname urlstyle\endcsname\relax
  \providecommand{\doi}[1]{doi:\discretionary{}{}{}#1}\else
  \providecommand{\doi}{doi:\discretionary{}{}{}\begingroup
  \urlstyle{rm}\Url}\fi
\providecommand{\eprint}[2][]{\url{#2}}

\bibitem[{Adibuzzaman \emph{et~al.}(2016)Adibuzzaman, Musselman, Johnson,
  Brown, Pitluk, and Grama}]{adibuzzaman2016}
Adibuzzaman M, Musselman K, Johnson A, Brown P, Pitluk Z, Grama A (2016).
\newblock \enquote{Closing the data loop: An integrated open access analysis
  platform for the mimic database.}
\newblock In \emph{2016 Computing in Cardiology Conference (CinC)}, pp.
  137--140. IEEE.

\bibitem[{Bennett(2021)}]{bennett2021}
Bennett N (2021).
\newblock \emph{prt: Tabular Data Backed by Partitioned 'fst' Files}.
\newblock R package version 0.1.3,
  \urlprefix\url{https://CRAN.R-project.org/package=prt}.

\bibitem[{Bone \emph{et~al.}(1992)Bone, Sibbald, and Sprung}]{bone1992}
Bone RC, Sibbald WJ, Sprung CL (1992).
\newblock \enquote{The ACCP-SCCM Consensus Conference on Sepsis and Organ
  Failure.}
\newblock \emph{Chest}, \textbf{101}(6), 1481 -- 1483.
\newblock ISSN 0012-3692.

\bibitem[{Desautels \emph{et~al.}(2016)Desautels, Calvert, Hoffman, Jay, Kerem,
  Shieh, Shimabukuro, Chettipally, Feldman, Barton
  \emph{et~al.}}]{desautels2016}
Desautels T, Calvert J, Hoffman J, Jay M, Kerem Y, Shieh L, Shimabukuro D,
  Chettipally U, Feldman MD, Barton C, \emph{et~al.} (2016).
\newblock \enquote{Prediction of sepsis in the intensive care unit with minimal
  electronic health record data: a machine learning approach.}
\newblock \emph{JMIR medical informatics}, \textbf{4}(3), e28.

\bibitem[{Evans(2016)}]{evans2016}
Evans R (2016).
\newblock \enquote{Electronic health records: then, now, and in the future.}
\newblock \emph{Yearbook of medical informatics}, \textbf{25}(S 01), S48--S61.

\bibitem[{Faltys \emph{et~al.}(2021)Faltys, Zimmermann, Lyu, Hüser, Hyland,
  Rätsch, and Merz}]{faltys2021}
Faltys M, Zimmermann M, Lyu X, Hüser M, Hyland SL, Rätsch G, Merz TM (2021).
\newblock \enquote{HiRID, a high time-resolution ICU dataset (version 1.1.1).}
\newblock PhysioNet.

\bibitem[{Fleuren \emph{et~al.}(2020)Fleuren, Klausch, Zwager, Schoonmade, Guo,
  Roggeveen, Swart, Girbes, Thoral, Ercole, Hoogendoorn, and
  Elbers}]{fleuren2019}
Fleuren LM, Klausch TLT, Zwager CL, Schoonmade LJ, Guo T, Roggeveen LF, Swart
  EL, Girbes ARJ, Thoral P, Ercole A, Hoogendoorn M, Elbers PWG (2020).
\newblock \enquote{{Machine learning for the prediction of sepsis: a systematic
  review and meta-analysis of diagnostic test accuracy}.}
\newblock \emph{Intensive Care Medicine}, \textbf{46}(3), 383--400.

\bibitem[{Futoma \emph{et~al.}(2017)Futoma, Hariharan, Sendak, Brajer, Clement,
  Bedoya, O'Brien, and Heller}]{futoma2017}
Futoma J, Hariharan S, Sendak M, Brajer N, Clement M, Bedoya A, O'Brien C,
  Heller K (2017).
\newblock \enquote{An improved multi-output gaussian process rnn with real-time
  validation for early sepsis detection.}
\newblock \emph{arXiv preprint arXiv:1708.05894}.

\bibitem[{Goldberger \emph{et~al.}(2000)Goldberger, Amaral, Glass, Hausdorff,
  Ivanov, Mark, Mietus, Moody, Peng, and Stanley}]{goldberger2000}
Goldberger AL, Amaral LAN, Glass L, Hausdorff JM, Ivanov PC, Mark RG, Mietus
  JE, Moody GB, Peng CK, Stanley HE (2000).
\newblock \enquote{PhysioBank, PhysioToolkit, and PhysioNet.}
\newblock \emph{Circulation}, \textbf{101}(23), e215--e220.

\bibitem[{Haas \emph{et~al.}(2016)Haas, Lange, Saugel, Petzoldt, Fuhrmann,
  Metschke, and Kluge}]{haas2016}
Haas SA, Lange T, Saugel B, Petzoldt M, Fuhrmann V, Metschke M, Kluge S (2016).
\newblock \enquote{Severe hyperlactatemia, lactate clearance and mortality in
  unselected critically ill patients.}
\newblock \emph{Intensive care medicine}, \textbf{42}(2), 202--210.

\bibitem[{Henry and Wickham(2020)}]{wickham2020}
Henry L, Wickham H (2020).
\newblock \emph{rlang: Functions for Base Types and Core R and 'Tidyverse'
  Features}.
\newblock R package version 0.4.9,
  \urlprefix\url{https://CRAN.R-project.org/package=rlang}.

\bibitem[{Hyland \emph{et~al.}(2020)Hyland, Faltys, H{\"{u}}ser, Lyu, Gumbsch,
  Esteban, Bock, Horn, Moor, Rieck, Zimmermann, Bodenham, Borgwardt,
  R{\"{a}}tsch, and Merz}]{hyland2020}
Hyland SL, Faltys M, H{\"{u}}ser M, Lyu X, Gumbsch T, Esteban C, Bock C, Horn
  M, Moor M, Rieck B, Zimmermann M, Bodenham D, Borgwardt K, R{\"{a}}tsch G,
  Merz TM (2020).
\newblock \enquote{{Early prediction of circulatory failure in the intensive
  care unit using machine learning}.}
\newblock \emph{Nature Medicine}, \textbf{26}(3), 364--373.

\bibitem[{Jiang \emph{et~al.}(2017)Jiang, Jiang, Zhi, Dong, Li, Ma, Wang, Dong,
  Shen, and Wang}]{jiang2017}
Jiang F, Jiang Y, Zhi H, Dong Y, Li H, Ma S, Wang Y, Dong Q, Shen H, Wang Y
  (2017).
\newblock \enquote{Artificial intelligence in healthcare: past, present and
  future.}
\newblock \emph{Stroke and vascular neurology}, \textbf{2}(4), 230--243.

\bibitem[{Johnson \emph{et~al.}(2016)Johnson, Pollard, Shen, Li-wei, Feng,
  Ghassemi, Moody, Szolovits, Celi, and Mark}]{johnson2016}
Johnson AE, Pollard TJ, Shen L, Li-wei HL, Feng M, Ghassemi M, Moody B,
  Szolovits P, Celi LA, Mark RG (2016).
\newblock \enquote{MIMIC-III, a freely accessible critical care database.}
\newblock \emph{Scientific data}, \textbf{3}, 160035.

\bibitem[{Johnson \emph{et~al.}(2018)Johnson, Aboab, Raffa, Pollard,
  Deliberato, Celi, and Stone}]{johnson2018}
Johnson AEW, Aboab J, Raffa JD, Pollard TJ, Deliberato RO, Celi LA, Stone DJ
  (2018).
\newblock \enquote{A Comparative Analysis of Sepsis Identification Methods in
  an Electronic Database.}
\newblock \emph{Critical care medicine}, \textbf{46}(4), 494--499.

\bibitem[{Jones(2012)}]{jones2012}
Jones M (2012).
\newblock \enquote{NEWSDIG: The National Early Warning Score Development and
  Implementation Group.}
\newblock \emph{Clinical Medicine}, \textbf{12}(6), 501--503.
\newblock \doi{10.7861/clinmedicine.12-6-501}.

\bibitem[{Kam and Kim(2017)}]{kam2017}
Kam HJ, Kim HY (2017).
\newblock \enquote{Learning representations for the early detection of sepsis
  with deep neural networks.}
\newblock \emph{Computers in biology and medicine}, \textbf{89}, 248--255.

\bibitem[{Klik(2020)}]{klik2020}
Klik M (2020).
\newblock \emph{fst: Lightning Fast Serialization of Data Frames}.
\newblock R package version 0.9.4,
  \urlprefix\url{https://CRAN.R-project.org/package=fst}.

\bibitem[{Lee \emph{et~al.}(2011)Lee, Scott, Villarroel, Clifford, Saeed, and
  Mark}]{lee2011}
Lee J, Scott D, Villarroel M, Clifford G, Saeed M, Mark R (2011).
\newblock \enquote{Open-access MIMIC-II database for intensive care research.}
\newblock In \emph{Annual International Conference of the IEEE Engineering in
  Medicine and Biology Society. IEEE Engineering in Medicine and Biology
  Society. Conference}, volume 2011, pp. 8315--8.

\bibitem[{{Moody} and {Mark}(1996)}]{moody1996}
{Moody} GB, {Mark} RG (1996).
\newblock \enquote{A database to support development and evaluation of
  intelligent intensive care monitoring.}
\newblock In \emph{Computers in Cardiology}, pp. 657--660.

\bibitem[{Nemati \emph{et~al.}(2018)Nemati, Holder, Razmi, Stanley, Clifford,
  and Buchman}]{nemati2018}
Nemati S, Holder A, Razmi F, Stanley MD, Clifford GD, Buchman TG (2018).
\newblock \enquote{An interpretable machine learning model for accurate
  prediction of sepsis in the ICU.}
\newblock \emph{Critical care medicine}, \textbf{46}(4), 547--553.

\bibitem[{Nichol \emph{et~al.}(2011)Nichol, Bailey, Egi, Pettila, French,
  Stachowski, Reade, Cooper, and Bellomo}]{nichol2011}
Nichol A, Bailey M, Egi M, Pettila V, French C, Stachowski E, Reade MC, Cooper
  DJ, Bellomo R (2011).
\newblock \enquote{Dynamic lactate indices as predictors of outcome in
  critically ill patients.}
\newblock \emph{Critical Care}, \textbf{15}(5), R242.

\bibitem[{Ooms(2014)}]{ooms2014}
Ooms J (2014).
\newblock \enquote{The jsonlite Package: A Practical and Consistent Mapping
  Between JSON Data and R Objects.}
\newblock \emph{arXiv preprint arXiv:1403.2805}.

\bibitem[{Pollard \emph{et~al.}(2018)Pollard, Johnson, Raffa, Celi, Mark, and
  Badawi}]{pollard2018}
Pollard TJ, Johnson AE, Raffa JD, Celi LA, Mark RG, Badawi O (2018).
\newblock \enquote{The eICU Collaborative Research Database, a freely available
  multi-center database for critical care research.}
\newblock \emph{Scientific data}, \textbf{5}, 180178.

\bibitem[{Singer \emph{et~al.}(2016)Singer, Deutschman, Seymour, Shankar-Hari,
  Annane, Bauer, Bellomo, Bernard, Chiche, Coopersmith, Hotchkiss, Levy,
  Marshall, Martin, Opal, Rubenfeld, van~der Poll, Vincent, and
  Angus}]{singer2016}
Singer M, Deutschman CS, Seymour CW, Shankar-Hari M, Annane D, Bauer M, Bellomo
  R, Bernard GR, Chiche JD, Coopersmith CM, Hotchkiss RS, Levy MM, Marshall JC,
  Martin GS, Opal SM, Rubenfeld GD, van~der Poll T, Vincent JL, Angus DC
  (2016).
\newblock \enquote{{The Third International Consensus Definitions for Sepsis
  and Septic Shock (Sepsis-3)}.}
\newblock \emph{JAMA}, \textbf{315}(8), 801--810.

\bibitem[{Subbe \emph{et~al.}(2001)Subbe, Kruger, Rutherford, and
  Gemmel}]{subbe2001}
Subbe C, Kruger M, Rutherford P, Gemmel L (2001).
\newblock \enquote{{Validation of a modified Early Warning Score in medical
  admissions}.}
\newblock \emph{QJM: An International Journal of Medicine}, \textbf{94}(10),
  521--526.

\bibitem[{Therneau and Grambsch(2000)}]{therneau2000}
Therneau TM, Grambsch PM (2000).
\newblock \emph{Modeling survival data: extending the Cox model}.
\newblock Springer, New York.

\bibitem[{Therneau and Lumley(2015)}]{therneau2015}
Therneau TM, Lumley T (2015).
\newblock \enquote{Package ‘survival’.}
\newblock \emph{R Top Doc}, \textbf{128}, 112.

\bibitem[{Thoral \emph{et~al.}(2021)Thoral, Peppink, Driessen, Sijbrands,
  Kompanje, Kaplan, Bailey, Kesecioglu, Cecconi, Churpek, Clermont, van~der
  Schaar, Ercole, Girbes, Elbers, Force, and the SCCM/ESICM Joint Data
  Science~Task}]{thoral2021}
Thoral PJ, Peppink JM, Driessen RH, Sijbrands EJG, Kompanje EJO, Kaplan L,
  Bailey H, Kesecioglu J, Cecconi M, Churpek M, Clermont G, van~der Schaar M,
  Ercole A, Girbes ARJ, Elbers PWG, Force obotAUMCDAC, the SCCM/ESICM Joint
  Data Science~Task (2021).
\newblock \enquote{{Sharing ICU Patient Data Responsibly Under the Society of
  Critical Care Medicine/European Society of Intensive Care Medicine Joint Data
  Science Collaboration: The Amsterdam University Medical Centers Database
  (AmsterdamUMCdb) Example}.}
\newblock \emph{Critical Care Medicine}, \textbf{Latest Articles}.

\bibitem[{Van~Beest \emph{et~al.}(2013)Van~Beest, Brander, Jansen, Rommes,
  Kuiper, and Spronk}]{van2013}
Van~Beest PA, Brander L, Jansen SP, Rommes JH, Kuiper MA, Spronk PE (2013).
\newblock \enquote{Cumulative lactate and hospital mortality in ICU patients.}
\newblock \emph{Annals of intensive care}, \textbf{3}(1), 6.

\bibitem[{Villar \emph{et~al.}(2013)Villar, P{\'{e}}rez-M{\'{e}}ndez, Blanco,
  A{\~{n}}{\'{o}}n, Blanch, Belda, Santos-Bouza, Fern{\'{a}}ndez, Kacmarek, and
  {Spanish Initiative for Epidemiology and Therapies for ARDS (SIESTA)
  Network}}]{villar2013}
Villar J, P{\'{e}}rez-M{\'{e}}ndez L, Blanco J, A{\~{n}}{\'{o}}n JM, Blanch L,
  Belda J, Santos-Bouza A, Fern{\'{a}}ndez RL, Kacmarek RM, {Spanish Initiative
  for Epidemiology and Therapies for ARDS (SIESTA) Network} S (2013).
\newblock \enquote{{A universal definition of ARDS: the PaO2/FiO2 ratio under a
  standard ventilatory setting—a prospective, multicenter validation study}.}
\newblock \emph{Intensive Care Medicine}, \textbf{39}(4), 583--592.

\bibitem[{Vincent \emph{et~al.}(1996)Vincent, Moreno, Takala, Willatts,
  De~Mendon{\c{c}}a, Bruining, Reinhart, Suter, and Thijs}]{vincent1996}
Vincent JL, Moreno R, Takala J, Willatts S, De~Mendon{\c{c}}a A, Bruining H,
  Reinhart C, Suter P, Thijs LG (1996).
\newblock \enquote{The SOFA (Sepsis-related Organ Failure Assessment) score to
  describe organ dysfunction/failure.}

\bibitem[{{Wang} \emph{et~al.}(2018){Wang}, {Sun}, {Schroeder}, {Ameko},
  {Moore}, and {Barnes}}]{wang2018}
{Wang} RZ, {Sun} CH, {Schroeder} PH, {Ameko} MK, {Moore} CC, {Barnes} LE
  (2018).
\newblock \enquote{Predictive Models of Sepsis in Adult ICU Patients.}
\newblock In \emph{2018 IEEE International Conference on Healthcare Informatics
  (ICHI)}, pp. 390--391.

\bibitem[{Wang \emph{et~al.}(2020)Wang, McDermott, Chauhan, Ghassemi, Hughes,
  and Naumann}]{wang2020}
Wang S, McDermott MB, Chauhan G, Ghassemi M, Hughes MC, Naumann T (2020).
\newblock \enquote{Mimic-extract: A data extraction, preprocessing, and
  representation pipeline for mimic-iii.}
\newblock In \emph{Proceedings of the ACM Conference on Health, Inference, and
  Learning}, pp. 222--235.

\bibitem[{Wickham and Hester(2020)}]{hester2020}
Wickham H, Hester J (2020).
\newblock \emph{readr: Read Rectangular Text Data}.
\newblock R package version 1.4.0,
  \urlprefix\url{https://CRAN.R-project.org/package=readr}.

\end{thebibliography}
\end{document}